%% file: main.tex
\documentclass[11pt,a4paper,bibtotoc,idxtotoc,headsepline,footsepline,footexclude,BCOR12mm,DIV13]{scrbook}

\input{components/info}

\input{components/settings}

\input{components/commands}

\makeglossary

\begin{document}

	\frontmatter

	\input{components/cover}
%

	\clearemptydoublepage
	
	\input{components/titlepage}

	\input{components/disclaimer}
	
	\input{components/acknowledgements}
	
	\input{components/abstract}

	\tableofcontents
  
  \input{components/outline}

  \input{components/acronyms}

	\mainmatter

		%
		%
		\part[Introduction and Theory]{Introduction and Theory}
		\label{part:introAndBackgroundTheory}
		\input{chapters/Introduction}

		%
		\part[Implementation]{Implementation}
		\label{part:secondP}
		\input{chapters/implementation}

		%
		\part[Quality Management]{Quality Management}
		\label{part:quality}
		\input{chapters/quality_management}
		
		%
		\part[Results]{Results}
		\label{part:results}
		\input{chapters/results}

		%
		%
		
		\part*{Appendix}
		\addcontentsline{toc}{part}{Appendix}
		
		\appendix 
		
		\input{chapters/oneAppendix}

  \clearemptydoublepage
  
	\bibliography{bibliography/literature}

\end{document}

%% file: components/info.tex
\def\doctype{Bachelorarbeit in Informatik}
\def\title{Implementation of an Android Framework for USB storage access without root rights}
\def\titleGer{Implementierung eines Android Frameworks f{\"u}r den Zugriff auf USB Speicher ohne Rootrechte}
\def\author{Magnus Jahnen}
\def\date{April 15, 2014}

\def\footertext{}

%% file: components/settings.tex
\renewcommand{\sectfont}{\normalfont \bfseries}        

\usepackage{scrpage2}
\ifoot[\footertext]{\footertext} 
\setkomafont{pagenumber}{\normalfont\rmfamily}

\usepackage{url}

\usepackage{listings}
\usepackage{color}

\definecolor{dkgreen}{rgb}{0,0.6,0}
\definecolor{gray}{rgb}{0.5,0.5,0.5}
\definecolor{pink}{rgb}{1,0.07,0.57}

\usepackage{acronym}

\usepackage{ifthen}

\usepackage{palatino}

\usepackage{pifont}

\usepackage{thumbpdf}

\usepackage{subfigure}

\usepackage{colortbl}

\usepackage{styles/tumlogo}

\usepackage{multirow}

\usepackage{color}

\usepackage{amsmath}
\usepackage{amsfonts}
\usepackage{amssymb}
\usepackage{textcomp}
\usepackage{yhmath} 

\usepackage{makeidx}

\usepackage{float}

\usepackage[german,english]{babel}
\usepackage[latin1]{inputenc}       

\usepackage{styles/shortoverview}

\ifx\pdftexversion\undefined
 \usepackage[dvips]{graphicx}
 \DeclareGraphicsExtensions{.eps,.epsi}
 \graphicspath{{figures/}{figures/review}} 
 \usepackage{rotating}
 \usepackage[hypertex,hyperindex=false,colorlinks=false]{hyperref}
\else 
 \usepackage[pdftex]{graphicx}
 \DeclareGraphicsExtensions{.jpg,.JPG,.png,.pdf,.eps,.pdf}
 \graphicspath{{figures/}} 
 
 \usepackage{float}
 
 \usepackage{rotating}
 \usepackage[pdftex,colorlinks=false,linkcolor=red,citecolor=red,%
 anchorcolor=red,urlcolor=red,bookmarks=true,%
 bookmarksopen=true,bookmarksopenlevel=0,plainpages=false%
 bookmarksnumbered=true,hyperindex=false,pdfstartview=%
 ]{hyperref}
\fi

%% file: components/commands.tex
\newcommand{\clearemptydoublepage}{%
  \ifthenelse{\boolean{@twoside}}{\newpage{\pagestyle{empty}\cleardoublepage}}%
  {\clearpage}}


%% file: components/cover.tex




\def\bcorcor{0.15cm}
\addtolength{\hoffset}{\bcorcor}

\thispagestyle{empty}

 \vspace{4cm}
\begin{center}
	       \oTUM{4cm}
	   
	   \vspace{5mm}     
	   \huge FAKULT{\"A}T F{\"U}R INFORMATIK\\ 
	   \vspace{0.5cm}
	 \large DER TECHNISCHEN UNIVERSIT{\"A}T M{\"U}NCHEN\\
    \vspace{1mm}
        
	\end{center}

\vspace{15mm}
\begin{center}

   {\Large \doctype}

  \vspace{20mm}
  
  {\huge\bf \title}\\

  \vspace{15mm}

  {\LARGE  \author}
  
  \vspace{10mm}
  
  \begin{figure}[h!]
  \centering
   \includegraphics[width=4cm]{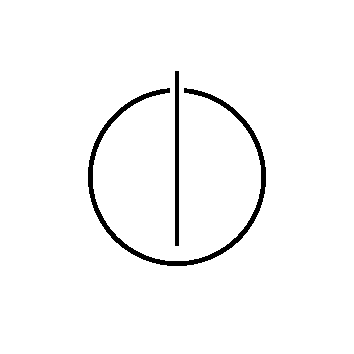}
  \end{figure}
  
  \end{center}

%% file: components/titlepage.tex


\def\bcorcor{0.15cm}
\addtolength{\hoffset}{\bcorcor}

\thispagestyle{empty}

 \vspace{10mm}
\begin{center}
	       \oTUM{4cm}
	   
	   \vspace{5mm}     
	   \huge FAKULT{\"A}T F{\"U}R INFORMATIK\\ 
	   \vspace{0.5cm}
	 \large DER TECHNISCHEN UNIVERSIT{\"A}T M{\"U}NCHEN\\
        
	\end{center}

\vspace{10mm}
\begin{center}

   {\Large \doctype}

  \vspace{10mm}
  
  {\LARGE \title}\\

  \vspace{10mm}

  {\LARGE  \titleGer}\\

  \vspace{10mm}

    \begin{tabular}{ll}
	   \Large Author:     & \Large \author \\[2mm]
	   \Large Supervisor:    & \Large Prof. Dr. Uwe Baumgarten \\[2mm]				
	   \Large Advisor:	& \Large Nils Kannengie{\ss}er, M.Sc. \\[2mm]
	   \Large Date:       & \Large April 15, 2014
	 \end{tabular}
	 
	 \vspace{5mm}
	 
	 \begin{figure}[h!]
  \centering
   \includegraphics[width=4cm]{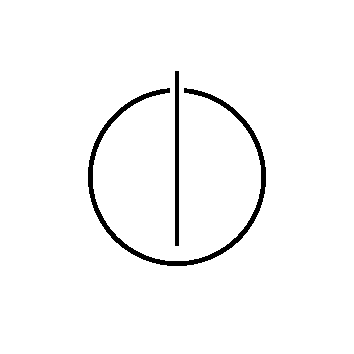}
  \end{figure}

\end{center}

\addtolength{\hoffset}{\bcorcor}

%% file: components/disclaimer.tex
\clearemptydoublepage

\thispagestyle{empty}
\selectlanguage{german}
	\vspace*{0.8\textheight}
	\noindent
	I assure the single handed composition of this bachelor thesis only supported by declared resources.
	
	\vspace{15mm}
	\noindent
	Munich, 15th of April, 2014 \hspace{7.5cm} \author
\selectlanguage{english}
\newpage

%% file: components/acknowledgements.tex
\clearemptydoublepage
\phantomsection
\addcontentsline{toc}{chapter}{Acknowledgements}


\vspace*{2cm}

\begin{center}
{\Large \bf Acknowledgments}
\end{center}

\vspace{1cm}

Many acknowledgment goes to the operating system chair of the TUM, especially to Prof. Dr. Uwe Baumgarten and Nils Kannengie{\ss}er for the ability to write this thesis and especially to Nils for guiding me. The operating system chair also borrowed me a lot of Android devices to test my work!

I also want to thank Jan Axelson for his book USB Mass Storage\cite{usb_ms_jan}. Without this book the thesis would have been a lot more cumbersome. It was an important resource throughout the whole work.

All people who read, corrected and gave me hints for improvement on this thesis in advance also deserve credit. I want to thank all of them at this place!

%% file: components/abstract.tex

\clearemptydoublepage
\phantomsection
\addcontentsline{toc}{chapter}{Abstract}

\vspace*{2cm}
\begin{center}
{\Large \bf Abstract}
\end{center}
\vspace{1cm}

This bachelor thesis describes the implementation of an Android framework to access mass storage devices over the USB interface of a smartphone. First the basics of USB (i.e. interfaces, endpoints and USB On the go) and accessing USB devices via the official Android API are discussed. Next the USB mass storage class is explained, which was designed by the USB-IF to access mobile mass storage like USB pen drives or external HDDs. For communication with mass storage devices, most important are the bulk-only transfer and the SCSI transparent command set. Furthermore file systems, for accessing directories and files, are described. This thesis focuses on the FAT32 file system from Microsoft, because it is the most commonly used file system on such devices.

After the theory part it is time to look at the implementation of the framework. In this section, the first concern is the purpose in general. Then the architecture of the framework and the actual implementation are presented. Important parts are discussed in detail.

The thesis finishes with an overview of the test results on various Android devices, a short conclusion and an outlook to future developments. Moreover the current status of the developed framework is visualized. 

%% file: components/outline.tex
\clearemptydoublepage

\phantomsection
\addcontentsline{toc}{chapter}{Outline of the Thesis}

\begin{center}
	\huge{Outline of the Thesis}
\end{center}

\section*{Part I: Introduction and Theory}

\noindent {\scshape Chapter 1: Introduction}  \vspace{1mm}

\noindent  Overview of the thesis and its purpose. \\

\noindent {\scshape Chapter 2: Basics about USB}  \vspace{1mm}

\noindent  Basics of USB, how it is structured and how it generally works.   \\

\noindent {\scshape Chapter 3: USB on Android}  \vspace{1mm}

\noindent  Introduction to the USB API on Android. How can USB devices be enumerated and accessed, and how can the communication be initiated? \\

\noindent {\scshape Chapter 4: USB Mass storage class}  \vspace{1mm}

\noindent  Overview of the USB mass storage class. The main goal is to understand SCSI commands to communicate with the device and read and write from and to its storage. \\

\noindent {\scshape Chapter 5: File systems}  \vspace{1mm}

\noindent  File systems in general and a detailed description of how FAT32 from Microsoft works. \\

\section*{Part II: Implementation}

\noindent {\scshape Chapter 6: Purpose and Overview}  \vspace{1mm}

\noindent  Purpose and requirements of the implementation and a short overview of the framework's architecture. \\

\noindent {\scshape Chapter 7: Inside the packages}  \vspace{1mm}

\noindent  Deeper insight into the developed framework, it's packages, classes and interaction between the important parts. \\

\section*{Part III: Quality Management}

\noindent {\scshape Chapter 8: Testing}  \vspace{1mm}

\noindent  Results of different tests regarding the developed framework on various devices. \\

\newpage

\section*{Part IV: Results}

\noindent {\scshape Chapter 9: Summary}  \vspace{1mm}

\noindent  Short summary of the current status of the framework and a conclusion of the thesis. \\

\noindent {\scshape Chapter 10: Outlook}  \vspace{1mm}

\noindent  Some hints about further development and advices what future may hold. \\

%% file: components/acronyms.tex
\clearemptydoublepage
\phantomsection
\addcontentsline{toc}{chapter}{Acronyms}

\vspace*{2cm}

\begin{center}
{\Large \bf Acronyms}
\end{center}

\vspace{1cm}

\begin{acronym}
\acro{adb}{Android Debug Bridge}
\acro{API}{Application Programming Interface}
\acro{app}{Application}
\acro{ASCII}{American Standard Code for Information Interchange}
\acro{ATAPI}{Advanced Technology Attachment with Packet Interface}
\acro{BBB}{bulk-only transport}
\acro{BIOS}{Basic input/output system}
\acro{btrfs}{B-tree file system}
\acro{CBI}{Control/Bulk/Interrupt}
\acro{CBS}{Command Status Wrapper}
\acro{CBW}{Command Block Wrapper}
\acro{CHS}{Cylinder-head-sector}
\acro{exFAT}{Extended File Allocation Table}
\acro{ext3}{Third extended file system}
\acro{ext4}{Fourth extended file system}
\acro{FAT}{File Allocation Table}
\acro{FS}{File System}
\acro{GPT}{GUID Partition Table}
\acro{GUID}{Globally Unique Identifier}
\acro{HDD}{Hard Disk Drive}
\acro{HFS+}{Hierarchical File System Plus}
\acro{ID}{Identifier}
\acro{IP}{Internet Protocol}
\acro{LBA}{Logical Block Address}
\acro{LFN}{Long File Name}
\acro{LUN}{Logical Unit Number}
\acro{MBR}{Master Boot Record}
\acro{NTFS}{New Technology File System}
\acro{OS}{Operating System}
\acro{RBC}{Reduced Block Commands}
\acro{ROM}{Read-only-memory}
\acro{SCSI}{Small Computer System Interface}
\acro{SD}{Secure Digital}
\acro{SPC}{SCSI Primary Commands}
\acro{UEFI}{Unified Extensible Firmware Interface}
\acro{UML}{Unified Modeling Language}
\acro{USB OTG}{USB On the go}
\acro{USB-IF}{USB Implementers Forum}
\acro{USB}{Universal Serial Bus}
\acro{xml}{Extensible markup language}
\end{acronym}

%% file: chapters/Introduction.tex
\chapter{Introduction}
\label{chapter:Introduction}

Since Android 3.1, which was originally designed for tablet computers, a lot of Android devices come with USB host support (USB On the go). That means a normal Android tablet or phone can not only act as an USB client when connected to a computer. It can also act as an USB host for peripherals by powering the bus with the needed five Volt and changing into USB host mode which enables enumeration of connected USB devices\cite{android_usb_host}. Android currently supports interrupt, bulk and control transfers\footnote{Isochronous transfers are currently unsupported\cite{android_usb_constants}.}. That means almost every USB device can, theoretically, be used with an Android device\footnote{Webcams or audio devices mostly use isochronous transfers and can thus not be used at the moment.}. The Android host API allows to communicate with connected USB devices, i.e. a high level USB driver can be written in Java.

Whereby the idea of connecting a USB mass storage device like USB flash drives or external HDDs is not far-fetched. Especially when looking at recent occurrences where a lot of devices lack a slot for SD-Cards and only offer a solid, mostly small, internal storage. Unfortunately the stock Android comes without support for USB storage devices. That means when connecting mass storage devices to an Android phone, nothing happens. The data cannot be accessed via a file manager or something similar. On rooted devices this is possible because the alternative Android ROMs often provide support for it. But with the Android USB Host API it should also be possible to access such devices without rooting the device and flashing an alternative ROM. The only thing needed is to implement the low level USB communication via e.g. bulk transfers and the abstraction of directories and files via a file system.

Currently there are two applications in the Google Play Store which allow accessing mass storage devices without root rights! First there is a plugin for the Total Commander called \textit{USB-Stick Plugin-TC}. The plugin extends the Total Commander application by USB mass storage access. It currently supports FAT12, FAT16, FAT32, exFAT and NTFS (read only). There is a free trial version available. The second application is called \textit{Nexus Media Importer}. It supports FAT16, FAT32 and NTFS (also read only). There is no free trial version available. In general both apps support USB sticks, external HDDs and card readers.

The disadvantage of both applications is that there is no solution to access the mass storage from other apps. That means all accessed data has to be cached and copied to the internal storage before any other app can use it. Unfortunately it seems that these limitations cannot be removed.

Both applications are proprietary and the source code is not available for reference or modification. This is why an open source Android framework for accessing mass storage devices is developed in this bachelor thesis. The desired license is the very liberal Apache License, Version 2.0, the same license Android is licensed under.

Due to the same licensing it would be possible for Google to integrate this solution into the official Android. But there are some factors which make the integration unlikely. First of all, necessary things, like file systems (e.g. FAT32) or the SCSI transparent command set, for mounting USB mass storage are already implemented in the underlying Linux kernel. Google just deactivated the support for it. Second, with the solution described in this thesis, only apps which use the framework can access USB storage devices. It would be more straightforward if the connected devices would be mounted in the normal unix file system like SD-cards. For example under \textit{/mnt/usbstick0}. This would allow other apps to easily access data from USB mass storage without extra changes to the application. Therefore it is very unlikely that Google will integrate this framework into the official Android. If Google decides to support mounting USB mass storage devices, they will most likely enable support for it in the kernel and mount the devices in the normal file system, like some manufacturers (e.g. Samsung) already do.

Another reason for Google not to implement the support for mass storage devices over USB, is to move more people to use their own cloud service Google Drive. But maybe, if Google notices the growing popularity of applications allowing access of USB mass storage, Google may enable support for it in their mobile operating system.

\subsubsection{Numbers in this thesis}

There are a lot of numbers in this thesis. If the number has a trailing 'h', for example 08h, this number shall be interpreted as a hex number. Numbers without this 'h' shall be interpreted as decimal numbers.

\chapter{Basics about USB}

USB stands for Universal Serial Bus and is a standard for a serial bus system, for connecting multiple peripheral devices to a personal computer. The first version was introduced by Intel in 1996. Today the specification is done by the USB Implementers Forum (USB-IF). The USB-IF is a corporation founded by various companies which work non-profit on the USB specification. In USB communication there are two kinds of devices, one USB host controller (e.g. computer) and one or more clients (slaves). The host is responsible for the communication process. The client only sends data when the host asks for it. The USB host is responsible for powering the connected client, thus an external power source is only necessary in some special cases\cite{wiki_usb}.

\section{Client device hierarchy}

A USB client is structured into four different USB descriptors:

\begin{itemize}
\item Device descriptor
\item Configuration descriptor
\item Interface descriptor
\item Endpoint descriptor
\end{itemize}

The device descriptor represents the USB device as a whole device which is connected to the USB bus. This can for example be a loud speaker with volume control buttons.

The configuration descriptor represents the current state of the USB device. This can for example be standby or active.

A USB interface descriptor describes every logical device which belongs to the USB device. Often USB devices consist of multiple logical device units. For example a loud speaker could consist of the speakers as an audio device and buttons to control the volume as a human interface device. 

Lastly there are endpoint descriptors which represent unidirectional communication pipes. This is where the actual communication happens. Endpoints can either be of type IN (device to host) or OUT (host to device). Additionally there are four different types of endpoints, to fit the different requirements of communication\cite{free_usb}.

\section{Endpoints}

Every USB device has different requirements on the underlying communication pipe. To satisfy the different needs the USB protocol offers four different types of communication (endpoints).

Control endpoints are used to configure the device and retrieve status information. Control transfers are typically very small. Every device has a control endpoint called endpoint 0 which plays an important role at insertion time\cite{free_usb}.

Interrupt transfers carry a small amount of data to the host every time the host asks for it. This happens at a fixed rate resulting in a fixed and guaranteed bandwidth. These transfers are used by human interface devices (e.g. mouse, keyboard, gamepads) which need a low latency and a low packet size.

Next there are bulk endpoints. These are useful when the amount of data transferred varies often and happens infrequently. The remaining available bandwidth the bus offers is used. Hence, there is no guarantee on bandwidth or latency. However bulk transfers offer consistent data transfers, meaning that no data is lost. They are typically used for printers, network or mass storage devices. Everywhere data loss is unacceptable and no guaranteed bandwidth is needed.

Finally there are the isochronous transfers. They offer a guaranteed bandwidth while resigning consistency. The guaranteed bandwidth is mostly as fast as possible and valuable for real time transfers (e.g. audio or video). Mostly these transfers are used for webcams or audio devices (e.g. external audio cards/audio interfaces)\cite{free_usb}.

\section{USB On the go}

As already mentioned, in USB communication there is always a host (master) and a client (device). The host initiates the communication and acts as a master. This means that the client can only send data after being explicitly asked to do so by the host. The client is only able to signal that it requires attention. Then the host must react and ask for receiving data. When connecting a smartphone or tablet to the computer, the computer acts as the host and the smartphone acts as the client device. That means the smartphone normally acts as a client device and not as the USB host. In the desired constellation described in this thesis, however, it has obviously to act as a host. 

For that reason the USB-IF developed the USB On the go (USB OTG) feature in 2001 as part of the USB 2.0 specification\cite{wiki_usb}. This feature allows a USB device to act as a client or either as a host depending on the present situation. To use the USB OTG mode, on a smartphone, a special USB OTG adapter is needed. This is necessary for two reasons. First for signaling that the smartphone should act as a host and not as usual as a client and second because most smartphones and tablets do not provide a normal USB port of type A. Instead they offer a mini (older devices) or micro port of type Mini-A or Micro-B\cite{wiki_usb_otg}. 

\chapter{USB on Android}
\label{chapter:usb_on_android}

As already mentioned, Google added USB features to the Android OS with Android 3.1 Honeycomb. There are two different modes Android keeps under control. The already mentioned host support and a special USB accessory mode. The accessory mode is only available on Android. It is supposed for developing USB host hardware specifically designed for Android devices, where the accessory is the USB host and powers the Android device\cite{android_usb_accessory}. The Android device is the client and can for example charge through and interact with the hardware (e.g. playing music through external speakers). 

The USB Accessory mode is also backported to Android 2.3.4\cite{android_usb_accessory}. The developed framework solely relies on the USB host functionality since a memory stick is a USB client.

\section{USB Host API}

Android offers classes to enumerate, connect to and communicate with connected USB devices. Table \ref{table:host_api} gives an overview of the classes which can be found in the package \textit{android.hardware.usb}.

\begin{table}[ht]
\caption{USB host APIs, compare to \cite{android_usb_host}}
\centering
\begin{tabular}{|l|p{10cm}|}
\hline\hline
\textbf{Class} & \textbf{Description} \\ \hline
UsbManager & Allows the enumeration and communication with connected USB devices. \\ \hline
UsbDevice & Represents a connected USB device and contains methods to access its identifying information, interfaces, and endpoints. \\ \hline
UsbInterface & Represents an interface of a USB device, which defines a set of functionality for the device. A device can have one or more interfaces. \\ \hline
UsbEndpoint & Represents an interface endpoint, which is a communication channel for this interface. An interface can have one or more endpoints, and usually has input and output endpoints for two-way communication with the device. \\ \hline
UsbDeviceConnection & Represents a connection to the device, which transfers data on endpoints. This class allows sending data back and forth synchronously or asynchronously. \\ \hline
UsbRequest & Represents an asynchronous request to communicate with a device through a UsbDeviceConnection. \\ \hline
UsbConstants & Defines USB constants that correspond to definitions in \textit{linux/usb/ch9.h} of the Linux kernel. \\ \hline
\end{tabular}
\label{table:host_api}
\end{table}

The UsbRequest class is only needed when communicating asynchronously\footnote{Asynchronous communication is passed in the implementation of the framework.}. The rough procedure of getting in touch with a USB device includes following steps:

\begin{enumerate}
\item Retrieve the desired UsbDevice via the UsbManager
\item Get the appropriate UsbInterface and UsbEndpoint
\item Begin the communication by opening a UsbDeviceConnection via the UsbEndpoint
\end{enumerate}

To understand the following sections, fundamental knowledge of Android programming is recommended. Basics\footnote{Following things, for example, are seen as basic: Activity, Intent, PendingIntent, IntentFiler, Broadcast, BroadcastReceiver.} are not described in detail here. An introduction to Android programming can be found in the official Android developer documentation\footnote{\url{http://developer.android.com}}.

\section{Enumerating devices}

To enumerate through all connected USB devices the singleton UsbManager is used. It allows looping through the device list. The device list is returned by the method getDeviceList() of the UsbManager.

\begin{lstlisting}[caption=Enumerating connected USB devices, label=listing:enumerate]
UsbManager usbManager = (UsbManager) context.getSystemService(Context.USB_SERVICE);
		
for(UsbDevice device : usbManager.getDeviceList().values()) {
	Log.i(TAG, "found usb device: " + device);
}
\end{lstlisting}

Accessing the UsbInterface and the UsbEndpoint is also straightforward. The UsbDevice has a method to get the desired interfaces and the UsbInterface has a method to get the desired endpoints on the other hand. Listing \ref{listing:interface_endpoint} illustrates the process.

\begin{lstlisting}[caption=Accessing UsbInterface and UsbEndpoint, label=listing:interface_endpoint]
UsbManager usbManager = (UsbManager) context.getSystemService(Context.USB_SERVICE);
		
for(UsbDevice device : usbManager.getDeviceList().values()) {
	Log.i(TAG, "found usb device: " + device);
	
	int interfaceCount = device.getInterfaceCount();
	for(int i = 0; i < interfaceCount; i++) {
		UsbInterface usbInterface = device.getInterface(i);
		Log.i(TAG, "found usb interface: " + usbInterface);
		
		int endpointCount = usbInterface.getEndpointCount();
		for(int j = 0; j < endpointCount; j++) {
			UsbEndpoint endpoint = usbInterface.getEndpoint(j);
			Log.i(TAG, "found usb endpoint: " + endpoint);
		}
	}
}
\end{lstlisting}

While looping through the devices, it can easily be checked if any device fits the desired needs. UsbDevice and UsbInterface offer methods to get the class, subclass and the protocol of the device, resp. the interface. The UsbEndpoint class has methods to get the type, the direction and other attributes of the corresponding endpoint. The UsbDevice also offers methods to get the vendor and product ID of the connected device.

\section{Requesting permission for communication}
\label{section:request_permission}

After discovering a suitable USB device Android requires the user to accept the communication between an application and the USB device first. To do so the permission to communicate with the USB device has to explicitly be requested. A dialog is shown to the user asking for permission, where the user can click okay or cancel. Therefore the UsbManager offers a method called requestPermission which takes a UsbDevice and a PendingIntent as parameter. The PendingIntent in this case is a Broadcast which can be received via registering a BroadcastReceiver with a specific IntentFilter.

Listing \ref{listing:permission_broadcast} shows how a BroadcastReceiver, for receiving notifications about the permission, can look like. First the intent action is validated, this step is only needed if the BroadcastReceiver receives multiple different actions. In this example this is not the case. After that the UsbDevice can be accessed via an extra of the intent. Another extra of the intent is the permission state. If it is granted the extra is true and it is permitted to communicate with the device.

\begin{lstlisting}[caption=Permission BroadcastReceiver \cite{android_usb_host}, label=listing:permission_broadcast]
private static final String ACTION_USB_PERMISSION =
    "com.android.example.USB_PERMISSION";
private final BroadcastReceiver mUsbReceiver = new BroadcastReceiver() {

    public void onReceive(Context context, Intent intent) {
        String action = intent.getAction();
        if(ACTION_USB_PERMISSION.equals(action)) {
            synchronized (this) {
                UsbDevice device = (UsbDevice)intent.getParcelableExtra(UsbManager.EXTRA_DEVICE);

                if(intent.getBooleanExtra(UsbManager.EXTRA_PERMISSION_GRANTED, false)) {
                    if(device != null){
                      //call method to set up device communication
                   }
                } 
                else {
                    Log.d(TAG, "permission denied for device " + device);
                }
            }
        }
    }
};
\end{lstlisting}

The next step is to register this BroadcastReceiver that it can actually receive broadcasts from the system. This normally happens in the onCreate method of an Activity via the method registerReceiver which takes the BroadcastReceiver (mUsbReceiver) and the IntentFilter as parameter. The IntentFilter uses the ACTION\_USB\_PERMISSION string, declared in Listing \ref{listing:permission_broadcast}, to filter undesired actions.

\begin{lstlisting}[caption=Registering the BroadcastReceiver \cite{android_usb_host}, label=listing:register_broadcast]
IntentFilter filter = new IntentFilter(ACTION_USB_PERMISSION);
registerReceiver(mUsbReceiver, filter);
\end{lstlisting}

The last step consists of requesting the permission using the UsbManager:

\begin{lstlisting}[caption=Requesting permission \cite{android_usb_host}, label=listing:request_permission]
UsbDevice device;
...
mPermissionIntent = PendingIntent.getBroadcast(this, 0, new Intent(ACTION_USB_PERMISSION), 0);
mUsbManager.requestPermission(device, mPermissionIntent);
\end{lstlisting}

\section{Communication}

After succeeding all necessary steps to set up the device, communication is possible. To do so the desired device has to be opened and an UsbDeviceConnection has to be retrieved. The class UsbDeviceConnection offers a method to claim a certain UsbInterface. After that communication is possible via the UsbDeviceConnection. It offers methods like bulkTransfer or controlTransfer.

\begin{lstlisting}[caption=Communicating with a connected device \cite{android_usb_host}, label=listing:communication]
private Byte[] bytes
private static int TIMEOUT = 0;
private boolean forceClaim = true;
...
UsbInterface intf = device.getInterface(0);
UsbEndpoint endpoint = intf.getEndpoint(0);
UsbDeviceConnection connection = mUsbManager.openDevice(device); 
connection.claimInterface(intf, forceClaim);
connection.bulkTransfer(endpoint, bytes, bytes.length, TIMEOUT); //do in another thread
\end{lstlisting}

Listing \ref{listing:communication} uses, for simplicity reasons, the first interface and endpoint. Normally the endpoint to communicate with, should be chosen wisely by examining for example the interface class, or the vendor ID of the device. A control transfer would look similar.

\section{Tearing down the communication}

When the communication between the Android application and the USB device is done, it has to be shut down. This is done by releasing the interface and closing the connection. Listing \ref{listing:closing_communication} gives an example about how to do that.

\begin{lstlisting}[caption=Closing communication, label=listing:closing_communication]
public void close() {
	Log.d(TAG, "close device");
	boolean release = deviceConnection.releaseInterface(usbInterface);
	if(!release) {
		Log.e(TAG, "could not release interface!");
	}
	deviceConnection.close();
}
\end{lstlisting}

\section{Listening to attach and detach events}

Android does not only allow enumerating connected devices, an application can also register for attach and detach events of USB devices. The application then gets notified whenever a USB device is connected to, or disconnected from the Android device. There are two different ways to do that. The first one is via a BroadcastReceiver, the second one via the AndroidManifest.xml file. The second method has the advantage that the application is notified even if it has not been started before.

\subsection{Via BroadcastReceiver}

\begin{lstlisting}[caption=Attach and detach notification of USB devices via a BroadcastReceiver, label=listing:attach_broadcast]
BroadcastReceiver mUsbReceiver = new BroadcastReceiver() {
    public void onReceive(Context context, Intent intent) {
        String action = intent.getAction(); 

      if (UsbManager.ACTION_USB_DEVICE_ATTACHED.equals(action)) {
            UsbDevice device = (UsbDevice)intent.getParcelableExtra(UsbManager.EXTRA_DEVICE);
            if (device != null) {
                // call method that sets up and initiates communication with the device
            }
        }

      if (UsbManager.ACTION_USB_DEVICE_DETACHED.equals(action)) {
            UsbDevice device = (UsbDevice)intent.getParcelableExtra(UsbManager.EXTRA_DEVICE);
            if (device != null) {
                // call method that cleans up and closes communication with the device
            }
        }
    }
};
\end{lstlisting}

To use this BroadcastReceiver it has to be registered in an Activity or Service with the corresponding IntentFilter like this:

\begin{lstlisting}[caption=Registering the BroadcastReceiver with the desired actions, label=listing:attach_register]
IntentFilter filter = new IntentFilter();
filter.addAction(UsbManager.ACTION_USB_DEVICE_ATTACHED);
filter.addAction(UsbManager.ACTION_USB_DEVICE_DETACHED);
registerReceiver(mUsbReceiver, filter);
\end{lstlisting}

\subsection{Via AndroidManifest.xml}

If an application wants to be notified about the attachment of an USB device this can also be specified in the AndroidManifest.xml. This has the advantage that the application does not have to be started before. In fact it is started when a desired USB device is connected. The user is then asked if he wants to start the application which can handle the attached device. The next benefit is that the step of requesting permission, described in \ref{section:request_permission}, is not required because the user already gave his consent by allowing the application to start.

Additionally a device filter can be specified, which allows the application to be notified only if an appropriate device is attached. Following attributes can be specified\cite{android_usb_host}:

\begin{itemize}
\item Vendor ID
\item Product ID
\item Class
\item Subclass
\item Protocol (device or interface)
\end{itemize}

Below is an example how a device filter could look like:

\lstset{language=XML}
\begin{lstlisting}[caption=Example device filter \cite{android_usb_host}, label=listing:device_filter]
<?xml version="1.0" encoding="utf-8"?>

<resources>
    <usb-device vendor-id="1234" product-id="5678" class="255" subclass="66" protocol="1" />
</resources>
\end{lstlisting}

This resource file should be located at \textit{res/xml/device\_filter.xml} in the project directory\cite{android_usb_host}. The device filter can then be used in the AndroidManifest.xml, like in Listing \ref{listing:manifest}.

\begin{lstlisting}[caption=AndroidManifest.xml \cite{android_usb_host}, label=listing:manifest]
<manifest ...>
    <uses-feature android:name="android.hardware.usb.host" />
    <uses-sdk android:minSdkVersion="12" />
    ...
    <application>
        <activity ...>
            ...
            <intent-filter>
                <action android:name="android.hardware.usb.action.USB_DEVICE_ATTACHED" />
            </intent-filter>

            <meta-data android:name="android.hardware.usb.action.USB_DEVICE_ATTACHED"
                android:resource="@xml/device_filter" />
        </activity>
    </application>
</manifest>
\end{lstlisting}

The intent filter for the action uses, again, the USB\_DEVICE\_ATTACHED string, like when using a BroadcastReceiver. This time no broadcast is sent, but an activity is started. The manifest also contains a \textit{uses-feature} entry, because not all Android devices guarantee to support the USB host feature\cite{android_usb_host}. The minimum sdk version is set to 12 here, because on lower API levels the USB host API is not available. \\\\
After that the UsbDevice can be accessed anywhere within the Activity like this:

\lstset{language=Java}
\begin{lstlisting}[caption={Accessing the UsbDevice in the Activity, compare to: \cite{android_usb_host}}, label=listing:access_usb_dev_activity]
UsbDevice device = (UsbDevice) getIntent().getParcelableExtra(UsbManager.EXTRA_DEVICE);
\end{lstlisting}

\chapter{USB Mass storage class}

Most USB devices are of same type. To reduce development effort and allow OS designers offering generic drivers for a great range of different devices, a lot of device types are standardized. These different types are called classes in USB. There are for example standardizations for printers, USB hubs, audio or video devices, human interface devices and mass storage devices\cite{usb_classes}. The focus in the following, is on the mass storage class.

Every mass storage device has at least one interface descriptor with the class code 08h, which stands for the mass storage class. The mass storage class is not defined in the device descriptor! The USB interface has exactly two endpoint descriptors. One IN endpoint to read from the device and one OUT endpoint to write to the device\cite{usb_ms_jan}. Reading and writing in this case does not necessarily mean reading or writing on the actual storage medium, this is described later.

There are two different types regarding the mass storage class. There is the bulk-only transport (BBB) mechanism which is the most common one. All newer devices follow that standard. Then there is the Control/Bulk/Interrupt (CBI) standard which is no longer important, because the USB-IF recommends using the BBB approach\cite{usb_ms_jan}.

\section{Bulk-only Transport}

Unlike the name suggests there are two control requests in the BBB specification. The first one is a reset request to prepare the device for the next command. The second is used to get the maximum LUN (Get Max LUN request). This request informs about the number of standalone logical units the mass storage device supports\cite{usb_ms_jan}.

As mentioned, the interface class has to be set to 08h for the mass storage class. The subclass of the interface descriptor can have different values and specifies the supported protocols used to read and write data from and to the mass storage. Table \ref{table:subclass} gives an overview of the different protocols.

\begin{table}[ht]
\caption{Overview subclass protocols \cite{usb_ms_jan}}
\centering
\begin{tabular}{|l|l|}
\hline\hline
01h & Reduced Block Commands (RBC) \\ \hline
02h & SFF-8020i, MMC-2 (ATAPI) (CD/DVD drives) \\ \hline
03h & QIC-157 (tape drives) \\ \hline
04h & USB Floppy Interface (UFI) \\ \hline
05h & FF-8070i (ATAPI removable rewritable media devices) \\ \hline
06h & SCSI transparent command set \\ \hline
\end{tabular}
\label{table:subclass}
\end{table}

For the purpose described in this thesis, the SCSI transparent command set is the most important one, which is explained in the following chapter. The RBC is not even implemented in Windows, but in the Linux kernel\cite{usb_ms_jan}. The other protocols refer to other types of storage media which are not covered by this thesis.

\section{SCSI transparent command set}

Every SCSI command the host sends to the client is enclosed by a so called Command Block Wrapper (CBW). Sending this CBW is always the first thing when host and device exchange data. After transmitting the CBW, raw data can be transferred. The direction of that data can either be from the host to the device or vice versa. In this document, from here on, this phase is called the data, transport or transfer phase. Some commands do not need the data phase. In the last step the client device sends a Command Status Wrapper (CSW) to the host to inform about any failures or success.

The CBW is always 31 bytes long including the enclosing SCSI command. The host sends it through the OUT endpoint to the device. Following table illustrates the CBW:

\begin{table}[ht]
\caption{Command Block Wrapper, compare to: \cite{usb_ms_jan}}
\centering
\begin{tabular}{|l|l|p{9cm}|}
\hline\hline
\textbf{Field Name} & \textbf{Bits} & \textbf{Description}\\ \hline
dCBWSignature & 32 & Fixed value of 43425355h to identify the CBW. \\ \hline
dCBWTag & 32 & Corresponds to dCSWTag in CSW. \\ \hline
dCBWDataTransferLength & 32 & The number of bytes which will be sent by the host in the transfer phase or the number of bytes the host expects to receive in the transfer phase. Depends on bmCBWFlags. \\ \hline
bmCBWFlags & 8 & If bit 7 is set to 0 the data transfer is from host to device, if it is set to 1 from device to host. All other bits are unused. If there is no transfer phase this value shall be zero. \\ \hline
Reserved & 4 & - \\ \hline
bCBWLUN & 4 & The LUN the command is directed to. \\ \hline
Reserved & 3 & - \\ \hline
bCBWCBLength & 5 & The length of the actual SCSI command located in the CBWCB field. \\ \hline
CBWCB & 128 & The SCSI command the client shall execute. \\ \hline
\end{tabular}
\label{table:cbw}
\end{table}

The dCBWTag is useful to associate the CSW with the CBW. The device uses the same value stored in the dCBWTag in the dCSWTag of the CSW. If multiple CBWs are sent at the same time the corresponding CSWs can easily be found with the help of the tag.
\\\\
What data is transferred and how, is discussed in the sections about the different SCSI commands. For now it is ignored and the CSW is introduced in table \ref{table:csw}. The CSW is always 13 bytes.

\begin{table}[ht]
\caption{Command Status Wrapper \cite{usb_ms_jan}}
\centering
\begin{tabular}{|l|l|p{9cm}|}
\hline\hline
\textbf{Field Name} & \textbf{Bits} & \textbf{Description}\\ \hline
dCSWSignature & 32 & Fixed value of 53425355h to identify the CSW. \\ \hline
dCSWTag & 32 & Value of dCBWTag from the CBW the device received. \\ \hline
dCSWDataResidue & 32 & This indicates the number of bytes in the transport phase the device has not yet processed. Should be 0 if all data has been processed. \\ \hline
bCSWStatus & 8 & 0 if command successfully passed, 1 if there was an error and 2 on a phase error. \\ \hline
\end{tabular}
\label{table:csw}
\end{table}

\newpage

The dCSWDataResidue holds the difference between the dCBWDataTransferLength and the number of bytes the device either processed when the host sends data or the number of bytes the device already sent to the host. In most cases all data can be transferred in one transfer phase meaning dCSWDataResidue is mostly zero.

The bCSWStatus informs about the success of executing the desired SCSI command. A value of zero indicates success. If this field is set to one there was an error executing the command. The host should then issue a SCSI REQUEST SENSE command to get more information about what went wrong\cite{usb_ms_jan}. More on this SCSI command later. If this value is two the host should perform a reset recovery. The reset consists of a bulk-only mass storage reset, which is one of the class specific commands and a Clear Feature HALT on the IN and OUT endpoint\cite{usb_ms_jan, usb_mass_bulk}.
\\\\
The fields in the CBW and CSW are all serialized in little endian style.

\section{SCSI commands}
\label{section:scsi_commands}

The Small Computer System Interface (SCSI) is a standard for communicating between computers and peripheral devices. It is most commonly used for hard disks and other storage devices, but it can also be used for example for scanners\cite{wiki_scsi}. SCSI commands are used to get general information about the connected storage device, but also for reading and writing data from and to the device's storage. The USB mass storage class also uses this well established standard.

There are a lot of different SCSI commands and not every device supports every command. To determine which commands are supported by a specific device, the host issues a SCSI INQUIRY command. Every device has to support this command and deliver a meaningful response to it. The device discloses, with the information included in the INQUIRY response, which commands are supported, i.e. which standardization it follows. In practice the most commonly supported commands are\cite{usb_ms_jan}:

\begin{itemize}
\item INQUIRY
\item READ CAPACITY(10)
\item READ(10)
\item REQUEST SENSE
\item TEST UNIT READY
\item WRITE(10)
\end{itemize}

Every device should support at least these commands! Every SCSI command starts with the operation code, also called OPCODE (one byte), which identifies the command. The following data depends on the specific command. The ten after some commands describes the length of the command in bytes. There are for example different READ commands, READ(6), READ(10), READ(12), READ(16) and READ(32)\cite{scsi_seagate}. These commands all differ in their length. This is needed because in the READ(6) command, the logical block address field which is used to address a block is only 16 bit. However devices with a big storage cannot use this command, because the whole storage cannot be addressed with a 16 byte value. Thus in the READ(10) command, which is the most commonly used read command, the address field is 32 bit.\\\\
SCSI commands use the big endian style for storing fields bigger than one byte.

\subsection{INQUIRY}

As already mentioned the INQUIRY command is used to get general information about the connected storage device. A host should issue this command to determine the supported SCSI commands by the device. The response to the INQUIRY command is transferred in the transport phase between sending the CBW which includes the INQUIRY command and receiving the CSW. The direction of the transport phase is from the client to the host.

\begin{table}[ht]
\caption{INQUIRY command, compare to: \cite{scsi_seagate}}
\centering
\begin{tabular}{|l|l|}
\hline\hline
\textbf{Byte} & \textbf{Description}\\ \hline
0 & Operation code (12h) \\ \hline
1 & Bit 0: EVPD, Bit 1: Obsolete, Bit 2-7: Reserved\\ \hline
2 & Page Code \\ \hline
3-4 & Allocation Length (Byte 3: MSB, Byte 4: LSB) \\ \hline
5 & Control \\ \hline
\end{tabular}
\label{table:inquiry}
\end{table}

The most important fields in the INQUIRY command are the operation code and the allocation length. The allocation length tells the storage device how many bytes the host has allocated for the INQUIRY response. The device then replies with an answer not larger than the allocation length. The Allocation Length field should be at least five (bytes). The EVPD\footnote{Enable Vital Product Data} and the page code are used to get more information about the vital product data. If the EVPD bit is set to one, the device should return the part of the vital product data specified in the field page code. If the EVPD bit is set to zero only the standard INQUIRY data shall be returned\cite{scsi_seagate}. This thesis describes only the latter case.

The Allocation Length field and the Control field are commonly used fields which occur in various SCSI commands\cite{scsi_seagate}.

The response to the standard INQUIRY request should contain at least 36 bytes\cite{scsi_seagate}. Nevertheless, it is up to the manufacturer of the device, how big the response is, because the response can include vendor specific information\cite{scsi_seagate}. Bytes 5 to N consist of fields not discussed here, because they are less important or vendor specific information.

\begin{table}[ht]
\caption{Standard INQUIRY data, compare to: \cite{usb_ms_jan, scsi_seagate}}
\centering
\begin{tabular}{|l|l|}
\hline\hline
\textbf{Byte} & \textbf{Description}\\ \hline
0 & Bit 0-4: Peripheral device type, Bit 5-7: Peripheral Qualifier \\ \hline
1 & Bit 0-6: Reserved, Bit 7: RMB \\ \hline
2 & Version \\ \hline
3 & Bit 0-3: Response data format, Bit 4: HISUP, Bit 5: NORMACA Bit 6,7: Obsolete \\ \hline
4 & Additional length (N-4) \\ \hline
5-N & ... \\ \hline
\end{tabular}
\label{table:inquiry_data}
\end{table}

The peripheral device type shall always be zero. This indicates that a peripheral device is connected to the logical unit. The peripheral qualifier describes the connected device. If this field is set to zero the connected device is a direct access block device. A value of two means a printer device and a value of five indicates a CD or DVD drive\cite{usb_ms_jan, scsi_seagate}. This value shall also always be zero for a direct access block device, because the direct acess block device is the only type of device the framework shall support.

The RMB bit indicates if the device is removable or not. Zero indicates a non removable device and one a removable device. USB flash drives are removable devices, but they have a fixed media unlike card readers. But Microsoft suggests that flash drives declare they have removable media, and thus some flash drives do this\cite{usb_ms_jan}.

The Version field indicates which standard of the SPC (SCSI Primary Commands) the device follows. If the value is zero the device does not comply to any standard. If the value is three or four, the device complies to the SPC or SPC-2 standard\cite{usb_ms_jan, scsi_seagate}.

The Response Data Format field must equal to two, because values lower than two are obsolete and values bigger than two are reserved\cite{scsi_seagate}.

The additional length provides information about how many bytes are remaining in the response. The additional data is not important at the moment.

\subsection{TEST UNIT READY}

This command tests if the storage device is ready to use. It does not have a transport phase. If the device is ready to use the CSW status is set to successful and if not to a status, indicating failure. In the latter case the host should issue a SCSI REQUEST SENSE, to get information about what went wrong. When the device has removable media, this command can be used to check if a media is currently present\cite{usb_ms_jan}.

\begin{table}[ht]
\caption{TEST UNIT READY command \cite{scsi_seagate}}
\centering
\begin{tabular}{|l|l|}
\hline\hline
\textbf{Byte} & \textbf{Description}\\ \hline
0 & Operation Code (00h)\\ \hline
1-4 & Reserved \\ \hline
5 & Control \\ \hline
\end{tabular}
\label{table:unit_ready}
\end{table}

\newpage

\subsection{READ CAPACITY}

The READ CAPACITY command is used to determine the storage space of a device. The device tells the host the logical block address (LBA) of the last block and the size in bytes of a single block. The total number of blocks is the LBA of the last block plus one. The direction of the transport phase is from the peripheral to the computer.

\begin{table}[ht]
\caption{READ CAPACITY(10) command \cite{scsi_seagate}}
\centering
\begin{tabular}{|l|l|}
\hline\hline
\textbf{Byte} & \textbf{Description}\\ \hline
0 & Operation Code (25h)\\ \hline
1 & Bit 0: Obsolete, Bit 1-7: Reserved \\ \hline
2-5 & Logical Block Address (Byte 2: MSB, Byte 5: LSB) \\ \hline
6,7 & Reserved \\ \hline
8 & Bit 0: PMI, Bit 1-7: Reserved \\ \hline
9 & Control \\ \hline
\end{tabular}
\label{table:read_capacity}
\end{table}

If the PMI (partial media indicator) bit is set to zero, the logical block address must also be set to zero. The device then returns information of the last logical block. If the PMI bit is set to one, the Seagate manual on SCSI commands says: "A PMI bit set to one specifies that the device server return information on the last logical block after that specified in the LOGICAL BLOCK ADDRESS field before a substantial vendor specific delay in data transfer may be encountered."\cite{scsi_seagate}

The response transferred in the transport phase looks like this:

\begin{table}[ht]
\caption{READ CAPACITY(10) response, compare to \cite{usb_ms_jan, scsi_seagate}}
\centering
\begin{tabular}{|l|l|}
\hline\hline
\textbf{Byte} & \textbf{Description}\\ \hline
0-3 & Last Logical Block Address (Byte 0: MSB, Byte 3: LSB)\\ \hline
4-7 & Block length in bytes (Byte 4: MSB, Byte 7: LSB) \\ \hline
\end{tabular}
\label{table:read_capacity_response}
\end{table}

\subsection{READ(10) and WRITE(10)}

The READ(10) command requests the device to read the specified blocks from the storage and to transfer them to the host. The logical block address included in the command specifies the block where reading shall begin. The Transfer Length field holds the amount of contiguous blocks that shall be read. The device then transmits the requested data in the data transport phase. The device does not have to care about the actual data, it transfers the data to the host, just like it is saved on the storage. Table \ref{table:read_10} shows how the READ(10) command is constructed.

\begin{table}[ht]
\caption{READ(10) command \cite{scsi_seagate}}
\centering
\begin{tabular}{|l|p{10cm}|}
\hline\hline
\textbf{Byte} & \textbf{Description}\\ \hline
0 & Operation Code (28h)\\ \hline
1 & Bit 0: Obsolete, Bit 1: FUA\_NV, Bit 2: Reserved, Bit 3: FUA, Bit 4: DPO, Bit 5-7: RDPROTECT \\ \hline
2-5 & Logical Block Address (LBA) (Byte 2: MSB, Byte 5: LSB) \\ \hline
6 & Bit 0-4: Group Number, Bit 5-7: Reserved \\ \hline
7,8 & Transfer Length (Byte 7: MSB, Byte 8: LSB) \\ \hline
9 & Control \\ \hline
\end{tabular}
\label{table:read_10}
\end{table}

For this thesis only the LBA and Transfer Length fields are important. The other fields shall remain zero. They are responsible, for example, to specify caching behavior or read protection\footnote{Way back, pen drives with a physical switch for write protection, were pretty common.}.

The WRITE(10) command is formatted exactly like the READ(10) command except that the operation code is 2Ah and the RDPROTECT field is called WDPROTECT. The direction of the transport phase is, of course, the other way round, from computer to the device. The direction has to be specified correctly in the CBW!

\subsubsection{Logical Block Address}

Every mass storage device is structured in blocks. These blocks have a defined size. The size of each block can be determined by issuing a READ CAPACITY(10) command. The blocks are numbered consecutively beginning from zero to the amount of blocks minus one. This number is called logical block address, short LBA. With the LBA every block can easily be addressed and accessed. The READ(10) and WRITE(10) SCSI commands use this method of addressing for reading and writing data from and to the storage medium. The transfer length specifies how many blocks, including the block at the LBA, shall be transferred. That means reading or writing begins with a block defined through the LBA with as many directly consecutive blocks as desired.

\subsection{REQUEST SENSE}

If the device fails executing a SCSI command requested by the computer, an unsuccessful CSW status is set. The computer then knows that something went wrong, but it does not know what went wrong. To get more information about a specific error the host can issue a REQUEST SENSE command, to request the sense data from the device.

\begin{table}[ht]
\caption{REQUEST SENSE command \cite{scsi_seagate}}
\centering
\begin{tabular}{|l|l|}
\hline\hline
\textbf{Byte} & \textbf{Description}\\ \hline
0 & Operation Code (03h)\\ \hline
1 & Bit 0: DESC, Bit 1-7: Reserved \\ \hline
2-3 & Reserved \\ \hline
4 & Allocation Length \\ \hline
5 & Control \\ \hline
\end{tabular}
\label{table:request_sense}
\end{table}

\newpage

The DESC bit describes if the fixed sense data or the descriptor format sense data shall be transferred. A value of zero requests the fixed sense data\cite{scsi_seagate}.

The Allocation Length field indicates, like in the INQUIRY command, how many bytes the host has allocated for data to be received. The device does not send more data than the host has actually allocated. But that means that some information can be lost if the data requested is actually bigger than the allocated space.

Table \ref{table:sense_data} shows the contents of the fixed sense data transferred from the device to the computer in the data transport phase. The size of the sense data normally is 252 bytes, with vendor specific information beginning at byte 18.

\begin{table}[ht]
\caption{Fixed SENSE data, compare to: \cite{usb_ms_jan, scsi_seagate}}
\centering
\begin{tabular}{|l|l|}
\hline\hline
\textbf{Byte} & \textbf{Description}\\ \hline
0 & Bit 0-6: Response Code, Bit 7: VALID \\ \hline
1 & Obsolete \\ \hline
2 & Bit 0-3: SENSE KEY, Bit 4: Reserved, Bit 5: ILI, Bit 6: EOM, Bit 7:  FILEMARK \\ \hline
3-6 & Information (Byte 3: MSB, Byte 6: LSB) \\ \hline
7 & Additional sense length (N-7) \\ \hline
8-11 & Command-specific information (Byte 8: MSB, Byte 11: LSB) \\ \hline
12 & Additional Sense Code \\ \hline
13 & Additional Sense Code Qualifier \\ \hline
14 & Field replaceable unit code \\ \hline
15-17 & Bit 0-20: Sense key Specific, Bit 21: SKSV \\ \hline
18-N & ... \\ \hline
\end{tabular}
\label{table:sense_data}
\end{table}

A detailed description of these fields can be found in the SCSI Commands Reference Manual from Seagate\cite{scsi_seagate}.

\chapter{File systems}

\section{General}

With the USB bulk transfers and the SCSI commands it is now possible to read and write raw data from and to the device. That means one can access the bytes stored on the medium. The only thing missing is some abstraction to handle the raw data. To do this on mass storage devices the most commonly used approaches are partitions and file systems. A file system is a way to organize data in directories and files with human readable names. Thus the user can easily find things without knowing something about the addressing methods of a mass storage device.

\subsubsection{Directories}

A directory is a container for files and other directories. With the help of a directory the contents of a file system or mass storage device can easily be structured in a tree based way. Every file system has a root directory which is the directory at the top of the file system.

\subsubsection{Files}

Unlike directories, files do not help structuring the contents, but hold the actual data the user wants to save and later access again.

\subsection{Examples}

Today there are many different file systems. Most commonly used file systems are FAT32 and NTFS from Windows background, ext3, ext4 and btrfs from the Linux/Unix background and HFS+, also called Mac OS Extended, developed by Apple for their OS X operating system. All listed file systems, except FAT32, use binary trees or something similar to structure the contents of the directories.

On USB mass storage devices the most popular file systems are FAT32 and NTFS, because obviously most people use the Windows operating system. The NTFS specification is not published by Microsoft. Thus it is very hard to support NTFS on other systems, nevertheless there is NTFS support in the Linux kernel. The FAT32 specification is publicly available and can be downloaded from the Microsoft website\footnote{\url{http://msdn.microsoft.com/en-us/windows/hardware/gg463080.aspx}}. Because of the lack of support for unix like file systems on Windows, the Linux and OS X file systems are only used by users who do not need to exchange data with Windows machines. Therefore the FAT32 file system, which has an open specification and is the mostly used file system on SD-cards and pen drives, is described in the following sections in detail.

\subsection{Partition table}

Before having a closer look at the FAT32 file system, partition tables have to be explained. A physical drive can have multiple partitions which operate independently from each other. They can also have different file systems. In Windows for example every partition is handled by a separate drive letter.

A partition table holds the information needed for identifying the different partitions on the disk. This information includes where the partition on the disk starts, ends or how many blocks it occupies and with which file system the partition is formatted. There are two different partition tables most commonly used today. The Master Boot Record (MBR) and the GUID partition table (GPT) which is part of the UEFI standard\cite{wiki_guid}. The Master Boot Record is used in PCs with BIOS and is currently replaced by the GPT in UEFI PCs. On USB mass storage devices mostly the MBR is applicable. 

\subsubsection{The Master Boot Record}

The MBR occupies 512 bytes at the beginning (LBA zero) of the medium. The first 446 bytes can store executable code the BIOS executes when a PC is booting. The executable code is then responsible for booting from an bootable partition. Beginning with byte 446 the partition table starts. There is place for up to four partition table entries. The last two bytes are the boot signature which identify the MBR. Byte 510 must be 55h and byte 511 must be AAh\cite{fat_paul}.

Since there is only place for four partition table entries, a disk formatted with the MBR normally could only have up to four different partitions, called primary partitions. If more partitions are required, extended partitions may be used. An extended partition has its own partition table enclosed by an extended boot record (EBR). The partition table entry in the MBR then points to the EBR which is followed by the extended partition. An EBR can contain one additional entry for another extended partition, thus there is practically no limit to the number of extended partitions\cite{usb_ms_jan}.

\begin{table}[ht]
\caption{Partition table entry in the MBR, compare to: \cite{usb_ms_jan}}
\centering
\begin{tabular}{|l|l|l|}
\hline\hline
\textbf{Byte} & \textbf{Size} & \textbf{Description}\\ \hline
0 & 1 & Boot Indicator, 00h for non bootable partition, 80h for bootable partition \\ \hline
1 & 1 & CHS addressing \\ \hline
2 & 2 & CHS addressing \\ \hline
4 & 1 & Partition Type \\ \hline
5 & 1 & CHS addressing \\ \hline
6 & 2 & CHS addressing \\ \hline
8 & 4 & LBA of the first sector \\ \hline
12 & 4 & Total number of sectors/blocks the partition occupies \\ \hline
\end{tabular}
\label{table:mbr_entry}
\end{table}

The important fields of a partition table entry in the MBR are located at byte four, eight and twelve. The fields for CHS addressing are pretty much obsolete. It is a former way to address the blocks on a block device through cylinder, head and sector information. Today only the logical block addressing mechanism is important\cite{wiki_lba}.

The partition table entry holds the first logical block address of the partition. This is where the partition starts, and the content of the file system of the desired partition begins. The entry also holds the total number of blocks the partition occupies, but in most cases this value is also stored in the file system of the partition, and most software uses this value instead\cite{usb_ms_jan}.

The partition type can either be a value indicating which file system the partition has, e.g. 0Bh or 0C for FAT32, or a hint for an extended partition. Extended partitions are identified by values of 05h or 0Fh. If the partition table entry is unused the partition type must equal to zero\cite{usb_ms_jan}.
\\\\
All fields bigger than one byte are stored using little endian style, as in the FAT32 file system.

\subsubsection{Drives without a partition table}

Sometimes drives do not have a partition table. If the device requires only one partition, the partition table is only a waste of space. If a device does not have a partition table, the file system directly begins at LBA zero.

But it is pretty hard to determine if the device has a partition table or if the file system directly starts, thus most devices have a partition table\cite{usb_ms_jan}. 

\section{The FAT32 file system}

The FAT32 file system was developed and published in 1996 with Windows 95. FAT means File Allocation Table, which is an important part of every FAT32 system, but more on that later. There are two ancestors of the FAT32 file system, FAT12 and FAT16. They vary in the size of an entry in the FAT and other fields. The entries of the FAT in a FAT32 file system have the size of 28 bit. Due to the 32 bit length field of a file, the file size is limited to 4 GiB - 1 Byte, which is sometimes unpleasant in everyday situations\cite{wiki_fat}.

\subsection{General layout}

The general layout of every FAT32 file system consists of following parts\cite{fatgen103, wiki_fat}:

\begin{enumerate}
\item Reserved Region (Boot Sector, FS Information Sector, optional reserved sectors)
\item File Allocation Table
\item Data area (directories and files)
\end{enumerate}

\begin{figure}[h!]
\caption{General layout of an FAT32 formatted volume}
\centering
\includegraphics[scale=0.62]{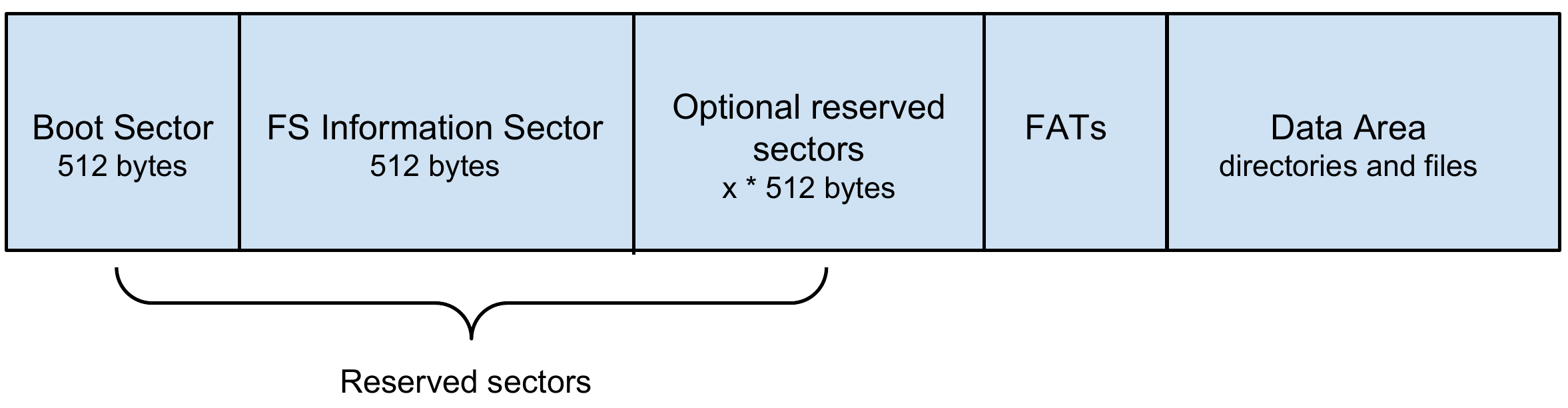}
\label{figure:general_fat32_layout}
\end{figure}

Figure \ref{figure:general_fat32_layout} illustrates the general layout from the beginning of a FAT32 formatted volume to its end. The sector size is assumed to be 512 bytes, which is mostly the case. Nevertheless, it can be different from 512 bytes. The sector size can be determined from the boot sector structure at the beginning of the medium. Directly after the reserved region the file allocation tables are located. The last and also biggest region is the data area. In the data area the actual content, like directories and files, is saved. Every part is described in detail in the following sections.

\subsubsection{Reserved Region}

The reserved region includes the boot sector and the FS Information sector. The boot sector holds important information of the FAT32 file system. For example the sector and the cluster size, the start cluster of the root directory and how many FATs exist. The FS Information Sector holds information about the last allocated cluster and the free cluster count. A file system driver can then easier locate a free cluster and give information about the remaining free space on the disk. After that optional reserved sectors can follow, for example a backup of the boot sector and the FS Information sector\cite{usb_ms_jan}.

\subsubsection{File Allocation Table}

The contents of the FAT32 file system is structured in so called clusters. A cluster has a specific size which is specified in the boot sector. A cluster is the smallest unit which can be allocated in the file system. That means if a file for example only has one byte of content but the cluster size is 4096 bytes the file needs 4096 bytes of space. The unneeded 4095 bytes are padded and wasted because they cannot be used for other files.

Every FAT32 file system has a definite amount of clusters which can be used for storing contents. The FAT gives an overview of which clusters are used and which are free. The FAT is a dynamically linked list. Giving a start cluster the FAT can be followed up to the end, identified by an end of chain mark. The resulting cluster chain helps locating the contents on the disk.

\subsubsection{Data area}

The data area holds the actual content, of the file system, the user wants to save. These are directories and files.

\subsection{Boot Sector and FS Information Structure}

\subsubsection{Boot Sector}

At the beginning of every FAT32 file system the boot sector is located. Table \ref{table:fat_boot_sector} shows the contents of the boot sector. The boot sector has the size of one sector (BPB\_BytsPerSec), which is typically 512 byte. Only the fields of interest are shown. A complete overview is given in the official FAT specification from Microsoft\cite{fatgen103}.

\begin{table}[!ht]
\caption{Boot Sector, compare to: \cite{usb_ms_jan, fatgen103}}
\centering
\begin{tabular}{|l|l|l|p{9cm}|}
\hline\hline
\textbf{Name} & \textbf{Offset} & \textbf{Size} & \textbf{Description}\\ \hline
BPB\_BytsPerSec & 11 & 2 & Count of bytes per a single sector. This is mostly 512 bytes, other allowed values are 1024, 2048 and 4096. A lot of file system drivers assume that this field is 512 and do not check it! \\ \hline
BPB\_SecPerClus & 13 & 1 & Count of sectors per cluster. This value must be a power of two greater than 0 and is mostly 8. The cluster size in bytes can be calculated with BPB\_BytsPerSec * BPB\_SecPerClus. \\ \hline
BPB\_RsvdSecCnt & 14 & 2 & Number of reserved sectors at the beginning of the volume, including the boot sector, preceding the FATs. This value is typically 32. \\ \hline
BPB\_NumFATs & 16 & 1 & The number of FATs in this file system. The FATs can be mirrored to provide redundancy to ensure that there is always a valid FAT which is not corrupt due to bad sectors or something else. This value is typically 2, meaning there are two different FATs holding the same information. \\ \hline
BPB\_TotSec32 & 32 & 4 & The total amount of sectors in the file system. \\ \hline
BPB\_FATSz32 & 36 & 4 & The number of sectors one FAT occupies. \\ \hline
BPB\_ExtFlags & 40 & 2 & Bit 0-3: Zero based number of the valid FAT if mirroring of FATs is disabled. Bit 7: Indicates if FATs are mirrored or not. 0 for mirroring, 1 if only one FAT is valid. Valid FAT can be determined with Bit 0-3. Other bits are reserved. \\ \hline
BPB\_RootClus & 44 & 4 & The start cluster of the root directory. This is typically cluster 2. \\ \hline
BPB\_FSInfo & 48 & 2 & The sector number of the FS Information Sector in the reserved region. \\ \hline
BPB\_BkBootSec & 50 & 2 & If this value is non zero, it indicates the sector number of the backup boot sector within the reserved region. This value is typically 6. \\ \hline
BS\_VolLab & 71 & 11 & Human readable string which gives the volume a name. This field is often replaced by a volume label entry in the root directory. \\ \hline
\end{tabular}
\label{table:fat_boot_sector}
\end{table}

The boot sector has the same boot signature at byte 510 and 511 like the MBR. That is another problem why it is problematic determining if the current drive has an MBR or if the file system starts directly.

\newpage

\subsubsection{FS Information Structure}

The FS Information Structure, also called FSInfo Sector Structure in Microsoft documents, helps finding free clusters quickly. Because the FAT can be very big in a FAT32 file system it can take a lot of time to go through the whole FAT to search for a free cluster. For that reason, the FSInfo Structure holds a hint to the last allocated cluster. It also stores the count of free clusters. The location of the sector is stored in the boot sector.

\begin{table}[!ht]
\caption{FSInfo Sector, compare to: \cite{fatgen103}}
\centering
\begin{tabular}{|l|l|l|p{9cm}|}
\hline\hline
\textbf{Name} & \textbf{Offset} & \textbf{Size} & \textbf{Description}\\ \hline
FSI\_LeadSig & 0 & 4 & Fixed value of 41615252h to identify the FSInfo Sector. \\ \hline
FSI\_Reserved1 & 4 & 480 & Reserved and normally set to zero. \\ \hline
FSI\_StrucSig & 484 & 4 & Fixed value of 61417272h to identify the FSInfo Sector. \\ \hline
FSI\_Free\_Count & 488 & 4 & The amount of free clusters in the volume. If FFFFFFFFh the free cluster count is unknown and should be computed. \\ \hline
FSI\_Nxt\_Free & 492 & 4 & The last known allocated cluster in the volume. A file system driver should use this to start searching for the free clusters. This does not mean the cluster after the last known is really free, it is just a hint! If FFFFFFFFh the last allocated cluster hint is unknown and the whole FAT has to be searched. \\ \hline
FSI\_Reserved2 & 496 & 12 & Reserved and normally set to zero. \\ \hline
FSI\_TrailSig & 508 & 4 & Fixed value of AA550000h to identify the FSInfo Sector. \\ \hline
\end{tabular}
\label{table:fat_fs_info}
\end{table}

\subsection{File Allocation Table}

As already mentioned, the File Allocation Table (FAT) is a dynamically linked list with a fixed size. It starts directly after the reserved region. The size of the FAT is stored in the boot sector of the FAT32 file system. The FAT is very important, because it holds the information about which clusters are used and which one not. Moreover it defines which different clusters combined define a file resp. directory. Thus the FAT can be mirrored, for backup reasons. The total count of FATs is located in the boot sector. The next FAT directly begins after the current FAT. Normally on a FAT32 file system there are two FATs holding the same information.

Every entry of the FAT is 32 bit, with the lower 28 bit representing the cluster number. The other four bits are unused. The first two entries in the FAT do not store any information about clusters, the data area begins with cluster 2\cite{usb_ms_jan}.

\subsubsection{Cluster chains}

A series of clusters is called a cluster chain. Every directory or file consists of a specific cluster chain defining the location of the content on the disk. To follow such a chain, a start cluster is needed. The start cluster of the root directory for example is stored in the boot sector. A FAT32 file system then seeks into the FAT where the start cluster is located. In bytes this is 4 * start cluster (Every entry is 32 bit = 4 byte) from the beginning of the FAT. The value stored at this location is the next cluster in the chain. This is repeated until the value stored is an "end of chain mark". This determines the end of a cluster chain. A value above FFFFFF7h is an end of chain mark. A value of zero indicates that the cluster is free and a value of one that the cluster is reserved. A value of FFFFFF7h stands for a bad cluster.

\begin{figure}[h!]
\caption{Simplified illustration of a FAT and following cluster chains}
\centering
\includegraphics{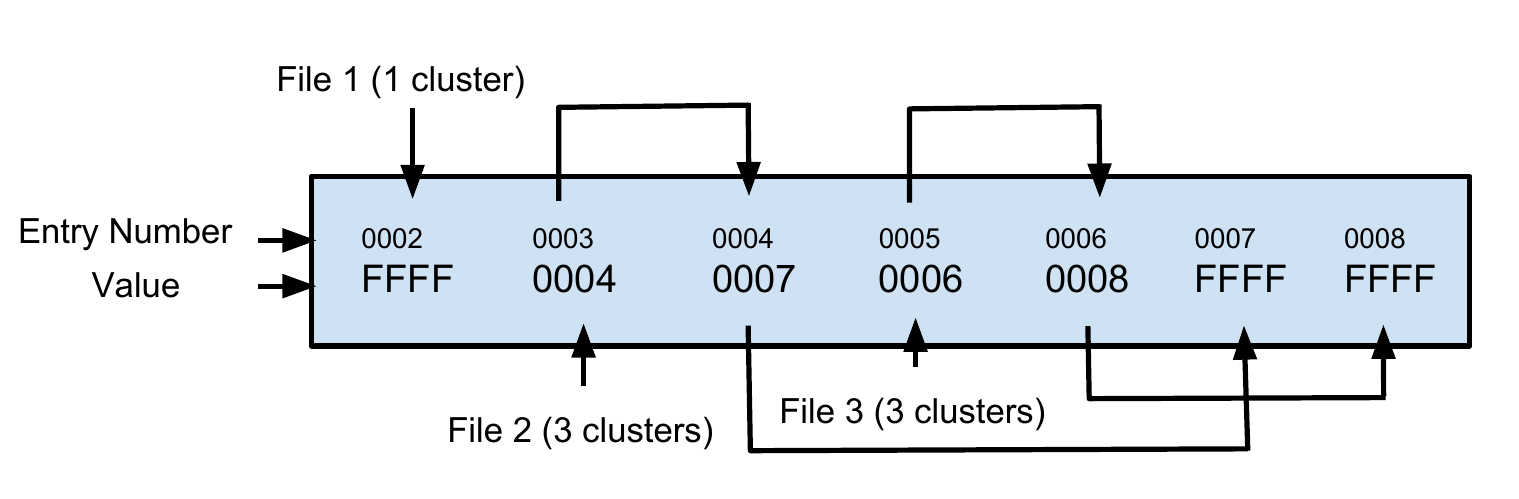}
\end{figure}

After evaluating the cluster chain, the file system driver can access the contents of the clusters in the data area. To do this, the logical block address where a cluster starts can be computed as follows: ((cluster - 2) * sectors per cluster) + data area offset

\subsection{Directories and files}

Directories and files are located in the data area of a FAT32 file system. The exact location of the contents and which clusters correspond to the directory or file must be determined by a cluster chain from the FAT. A directory consists of multiple 32 byte entries describing the contents of the directory. The contents can either be other directories, so called subdirectories or files. The root directory is always present on a FAT32 file system. The start cluster of the root directory is stored in the boot sector.

Files do not have a defined structure, unlike directories. The raw data the file consists of is located in the different clusters, just as it is. The only thing defined, is the order, which can be determined from the FAT.

\begin{figure}[h!]
\caption{Simplified illustration of the content in the data area}
\centering
\includegraphics[scale=0.62]{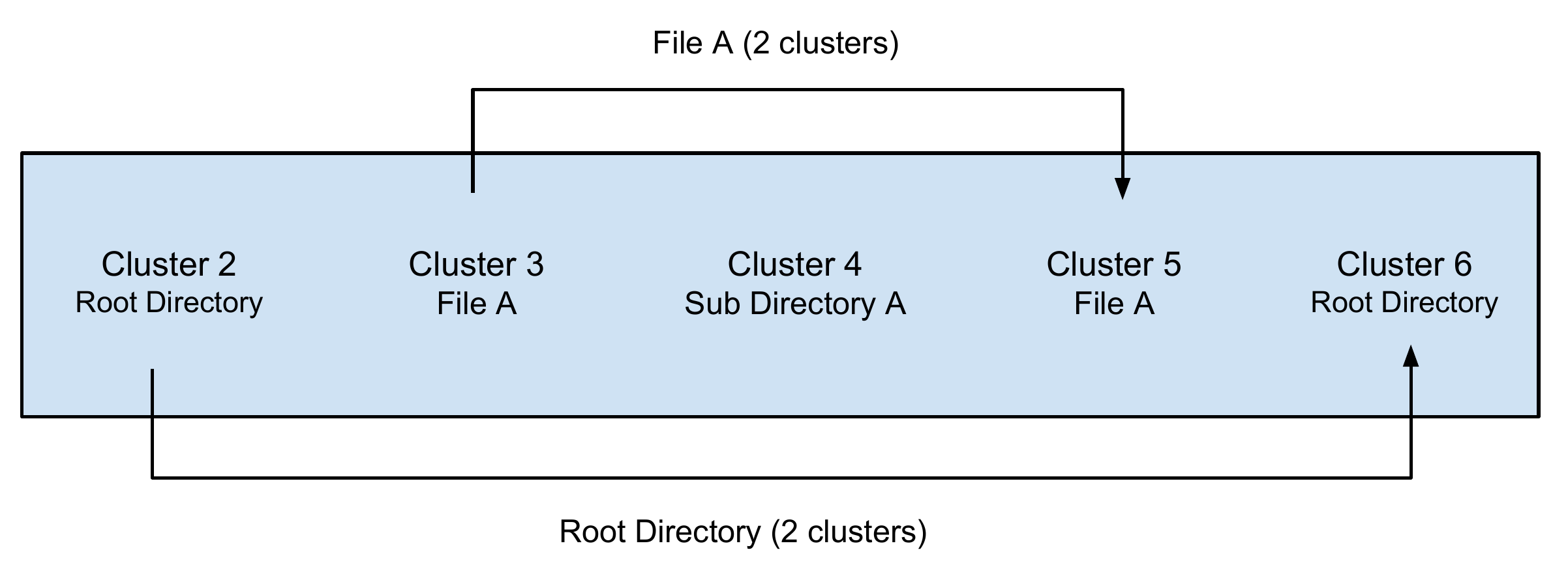}
\label{figure:data_area}
\end{figure}

Figure \ref{figure:data_area} shows a simplified example of the data area. The root directory starts at cluster 2 and ends with cluster 6. There is also a file in the data area, consisting of cluster 3 and 5. In cluster 4 a sub directory is located consisting of one cluster only. The whole data area is separated into clusters with a certain size, containing either chunks of data of a directory or a file.

\subsubsection{Fat Directory Entry}

The following table describes the structure of a Fat Directory entry. Every entry, normally, represents either another (sub)directory or file. The root directory can also have a special entry not only describing files and other directories but an optional volume label which gives the volume a name. Such a volume label could also be specified in the boot sector, but it is more common to specify it in the root directory. Another special entry is the long file name entry, which is described later.

\begin{table}[!ht]
\caption{Fat Directory Entry, compare to: \cite{usb_ms_jan, fatgen103}}
\centering
\begin{tabular}{|l|l|l|p{8.5cm}|}
\hline\hline
\textbf{Name} & \textbf{Offset} & \textbf{Size} & \textbf{Description}\\ \hline
DIR\_Name & 0 & 11 & The short name of the directory or file. \\ \hline
DIR\_Attr & 11 & 1 & Attributes to determine if this entry describes a directory or a file, etc. Described in table \ref{table:fat_dir_entry_attr} \\ \hline
DIR\_NTRes & 12 & 1 & Reserved. \\ \hline
DIR\_CrtTimeTenth & 13 & 1 & Timestamp which holds the tenth of a second when the file was created. The range is from 0 to 199. \\ \hline
DIR\_CrtTime & 14 & 2 & The time the file was created. \\ \hline
DIR\_CrtDate & 16 & 2 & The date the file was created. \\ \hline
DIR\_LstAccDate & 18 & 2 & The date the file was last accessed (read or write). \\ \hline
DIR\_FstClusHI & 20 & 2 & The higher two bytes of the entry's start cluster. Unused if the entry is not a file or directory. \\ \hline
DIR\_WrtTime & 22 & 2 & The time the file last modified (write). \\ \hline
DIR\_WrtDate & 24 & 2 & The date the file last modified (write). \\ \hline
DIR\_FstClusLO & 26 & 2 & The lower two bytes of the entry's start cluster. Unused if the entry is not a file or directory. \\ \hline
DIR\_FileSize & 26 & 4 & The length of a file in bytes. Unused if the entry is not a file. \\ \hline
\end{tabular}
\label{table:fat_dir_entry}
\end{table}

\subsubsection{Short name}

Every entry has a short name which is a human readable name. The name consists of eight characters for the file name and three optional characters for the file extension. Therefore it is also called 8.3 name. The period to separate name and extension is not stored in the short name. If a name does not need the whole eight characters the unused characters have to be padded with spaces (ASCII: 20h). The same applies for the extension. A short name has various limitations\cite{usb_ms_jan}:

\begin{itemize}
\item Each character is only eight bit (no unicode support)
\item It has to begin with a letter or a number
\item Every character is stored upper case
\item The only allowed special characters are: \$ \% $'$ -  \_ @ $\sim$ $`$ ! ( ) { } \^{} \# \&
\end{itemize}

If the first byte of the short name equals to E5h, then this directory entry is free and has been deleted. If the first byte equals to 00h, then this entry is also free, but there are not any following entries, looping through the entries can thus be stopped. In the first case it can happen that there are valid entries that follow. If the first byte equals to 05h, then the actual value of the first byte shall be E5h, which is a KANJI lead byte used in Japan. This is a workaround for avoiding the entry to be accidentally treated as free (deleted)\cite{fatgen103}.

\subsubsection{Attributes}

\begin{table}[!ht]
\caption{Attributes of an entry, compare to: \cite{usb_ms_jan, fatgen103}}
\centering
\begin{tabular}{|l|l|p{9cm}|}
\hline\hline
\textbf{Name} & \textbf{Value}  & \textbf{Description}\\ \hline
ATTR\_READ\_ONLY & 01h & The file is read only and writing to it should fail. \\ \hline
ATTR\_HIDDEN & 02h & A hidden entry the user should only see when explicitly asking for it. \\ \hline
ATTR\_SYSTEM & 04h & A file from the operating system. \\ \hline
ATTR\_VOLUME\_ID & 08h & This entry is not a file and not a directory, it is the entry which stores the volume label. This can occur only in the root directory and only once. \\ \hline
ATTR\_DIRECTORY & 10h & Indicates that the entry describes a (sub)directory. \\ \hline
ATTR\_ARCHIVE & 20h & Value helping backup utilities identifying files that changed since the last backup. Should be set if a file has been changed. \\ \hline
ATTR\_LONG\_NAME & 0fh & Indicates that this entry is not a file or a directory but an entry which holds a long file name. \\ \hline
\end{tabular}
\label{table:fat_dir_entry_attr}
\end{table}

The last attribute in table \ref{table:fat_dir_entry_attr} stands for a long file name entry. Because of the limitations of the short name, Microsoft added support for long file names afterwards. To ensure backward compatibility they used some sort of work around to hide the long file names in normal directory entries. They are discussed later in detail.

If the volume id attribute is set, the short name is not actually the name of a file, it is the name of the volume. The volume name is shown in the Windows Explorer directly left of the drive letter. The volume label can have up to eleven characters, but there is no period between the eighth and the ninth character\cite{usb_ms_jan, fatgen103}.

\subsubsection{Date and time fields}

In the directory entry there are many date and time fields, the following describes how the date and time is stored in these fields. Every date and time field consists of two bytes. 

\begin{table}[!ht]
\caption{Date format, compare to: \cite{usb_ms_jan, fatgen103}}
\centering
\begin{tabular}{|l|p{9cm}|}
\hline\hline
\textbf{Bits} & \textbf{Description}\\ \hline
0-4 & Day of month, range 1-31 \\ \hline
5-8 & Month, starting with 1 for January \\ \hline
9-15 & Count of years from 1980, range 0-127 corresponding to 1980-2107 \\ \hline
\end{tabular}
\label{table:fat_dir_entry_date}
\end{table}

\begin{table}[!ht]
\caption{Time format, compare to: \cite{usb_ms_jan, fatgen103}}
\centering
\begin{tabular}{|l|p{9cm}|}
\hline\hline
\textbf{Bits} & \textbf{Description} \\ \hline
0-4 & Count of seconds with a resolution of 2 seconds, range 0-29 meaning 0-58 seconds \\ \hline
5-10 & Minutes, range 0-59 \\ \hline
11-15 & Hours, range 0-23 \\ \hline
\end{tabular}
\label{table:fat_dir_entry_time}
\end{table}

\subsection{Subdirectories}

Every subdirectory has two special entries. The dot (.) and the dotdot (..) entry. The dot entry points the the current subdirectory itself. It has the same values as the entry for the subdirectory in the parent directory, i.e. same date and time fields, same start cluster, etc. The dotdot entry points to the parent directory, but the date and time fields remain like in the subdirectory. The start cluster of the dotdot entry is the same as for the parent directory except if the parent directory is the root directory, then it is set to zero\cite{fatgen103}.

These two entries are the only exceptions where a short name starts with periods. The dot and dotdot entry must not have preceding long file name entries\cite{fatgen103}!

The root directory does not have these two special entries, this rule applies for subdirectories only\cite{fatgen103}!

\subsection{Long File Name entries}

As already mentioned the short name of a directory entry has several disadvantages. For that reason Microsoft afterwards faced this problem by adding long file name (LFN) entries. These entries "hide" in the normal directory entries and look like a hidden file with a start cluster of zero, indicating that the file does not occupy any space. Thus old implementations which do not support long file name entries can nevertheless work with the file system and the 8.3 names.

Long file names can have up to 255 characters and allow upper and lower case characters, leading, trailing and multiple periods and spaces in a file name\cite{usb_ms_jan}. Additionally these special characters are allowed\cite{fatgen103}: + , ; = [ ] 

Each entry with a normal short name, can have one or more long file name entries preceding the actual entry. A long file name entry can store up to 13 unicode characters, i.e. every character consists of two bytes instead of eight like in the short name.

\begin{table}[!ht]
\caption{Fat LFN Directory Entry, compare to: \cite{usb_ms_jan, fatgen103}}
\centering
\begin{tabular}{|l|l|l|p{8.5cm}|}
\hline\hline
\textbf{Name} & \textbf{Offset} & \textbf{Size} & \textbf{Description}\\ \hline
LDIR\_Ord & 0 & 1 & The number (order) of the entry in the sequence of long file name entries. \\ \hline
LDIR\_Name1 & 1 & 10 & The first five characters of the long file name. \\ \hline
LDIR\_Attr & 11 & 1 & Must be ATTR\_LONG\_NAME. \\ \hline
LDIR\_Type & 12 & 1 & Must be zero. \\ \hline
LDIR\_Chksum & 13 & 1 & The checksum of the short name associated with this long file name. \\ \hline
LDIR\_Name2& 14 & 12 & The next six characters of the long file name. \\ \hline
LDIR\_FstClusLO & 26 & 2 & Must be zero. Long file name entries do not have a start cluster. \\ \hline
LDIR\_Name3I & 28 & 4 & The last two characters of the long file name. \\ \hline
\end{tabular}
\label{table:fat_lfn_dir_entry}
\end{table}

The first byte is the number of the LFN entry. The first LFN entry before the short name entry, must have a value of one in that field. The second a two, and so on. If the LFN entry is the last entry, bit six of the LDIR\_Ord field has to be set to one, to indicate that it is the last entry. There can be at most 20 LFN entries preceding a normal entry. The first LFN entry directly before the short name entry represents the begin of the long file name. That means the long file name is stored in "reverse order" on the disk. 

The long file name entry should be terminated with a null character (0h) if there is enough space, and unused characters shall be padded with FFFFh\cite{fatgen103}.

\subsubsection{Checksum}

Every long file name entry needs a checksum of the corresponding short name. This checksum is calculated as follows:

\lstset{language=c}
\begin{lstlisting}[caption=Calculation of short name checksum in C \cite{fatgen103}, label=listing:fat_checksum]
//------------------------------------------------------------------
// ChkSum()
// Returns an unsigned byte checksum computed on an unsigned byte
// array. The array must be 11 bytes long and is assumed to contain
// a name stored in the format of a MS-DOS directory entry.
// Passed: pFcbName Pointer to an unsigned byte array assumed to be 11 bytes long.
// Returns: Sum An 8-bit unsigned checksum of the array pointed to by pFcbName.
//-------------------------------------------------------------------
unsigned char ChkSum (unsigned char *pFcbName)
{
    short FcbNameLen;
    unsigned char Sum;
    
    Sum = 0;
    for (FcbNameLen=11; FcbNameLen!=0; FcbNameLen--) {
        // NOTE: The operation is an unsigned char rotate right
        Sum = ((Sum & 1) ? 0x80 : 0) + (Sum >> 1) + *pFcbName++;
    }
    return (Sum);
}
\end{lstlisting}

\subsubsection{Generating short names}

Every directory entry which has preceding long file name entries nevertheless needs a valid short name. There exist different algorithms for generating valid short names given a long file name. These algorithms mainly consist of removing illegal characters and convert them to underscores (\_), removing spaces and periods, truncating the long file name to eight and the extension, if available, to three characters. All characters must be converted to upper case and a tilde ($\sim$) must be added if the file name was truncated, contained illegal characters or an entry with the same short name already exists in the directory. The official FAT specification\cite{fatgen103} and the book from Jan Axelson\cite{usb_ms_jan} describe two different approaches to generate valid 8.3 short names.

%% file: chapters/implementation.tex
\chapter{Purpose and Overview}

\subsubsection{Purpose}
\label{implementation_purpose}

The developed Android framework for accessing USB mass storage devices shall meet determinate requirements:

\begin{itemize}
\item The mass storage device is accessible without root rights
\item The API provides methods for enumerating through all connected mass storage devices and their partitions
\item The API is easy to use and orientates towards the java.io.File API
\item The framework provides basic features like: Adding and removing directories and files, read and write access to files, moving directories and files located on the same volume
\item The framework shall be licensed under the Apache License, Version 2\footnote{\url{http://www.apache.org/licenses/LICENSE-2.0.html}}
\end{itemize}

The framework shall at least support \textit{bulk-only transport} mass storage devices, which are using the SCSI transparent command set. It shall support devices which are formatted with the MBR partition table and the FAT32 file system. Despite the framework a simple example application shall be developed also to test and demonstrate the framework. The framework and the example application are publicly available at github\footnote{\url{https://github.com/mjdev/libaums}}.

The following sections describe the relevant parts of the framework, but not every single detail. To get an insight how the whole thing works, the source code, which is well documented, may be examined directly.

\subsubsection{Overview}
\label{implementation_overview}

This sections gives a brief overview over the whole framework, in the following chapters important parts of the framework are discussed in detail.

The framework and the example application are written in Java. This is Android's main language for developing applications. The framework uses the standard Java API and the Android USB host API to access USB devices. The example application uses the API of the developed framework and the standard Android API for creating user interfaces.

\begin{figure}[h!]
\caption{Package overview of the framework}
\centering
\includegraphics[scale=0.9]{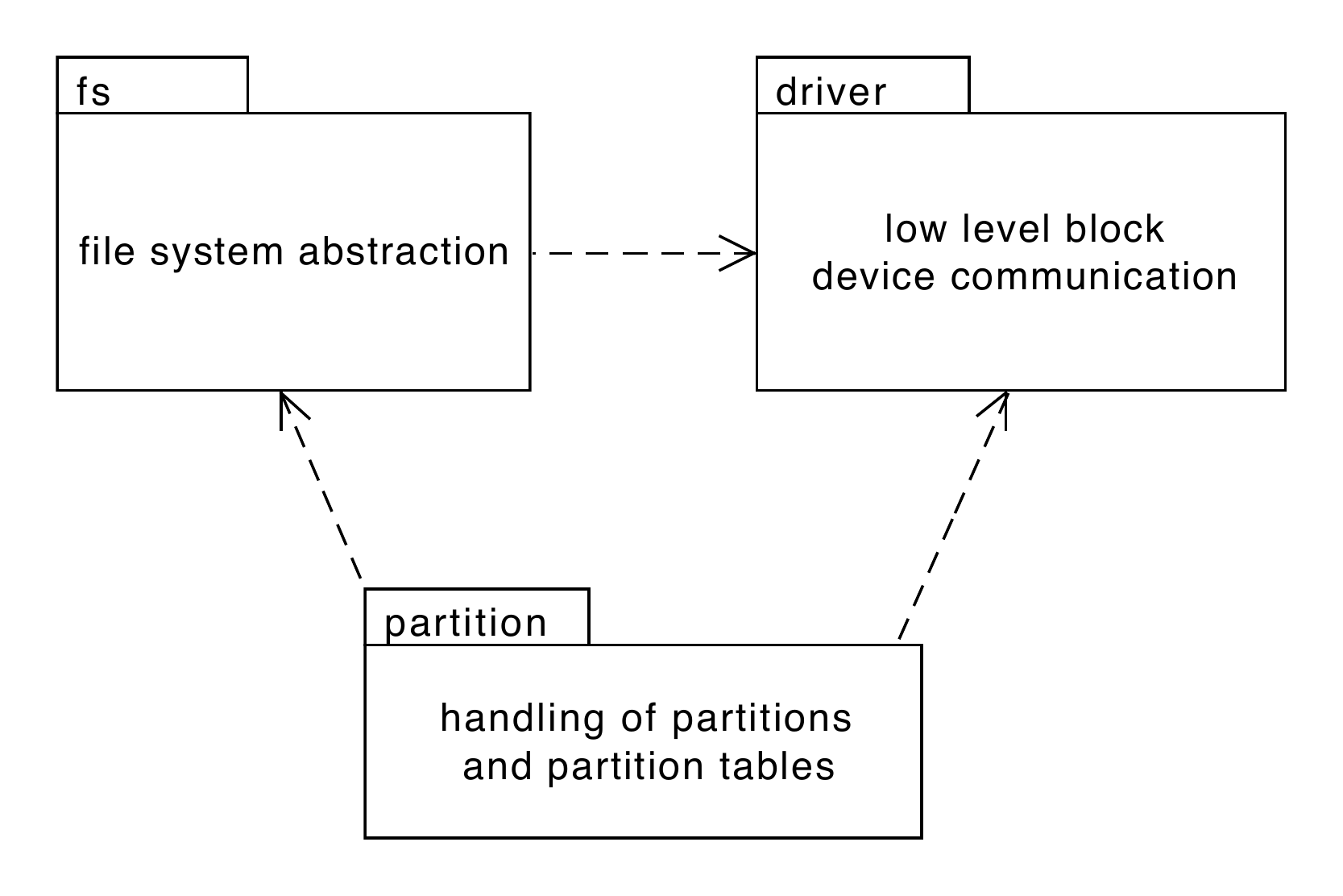}
\label{figure:package}
\end{figure}

The framework can roughly be structured in three parts, figure \ref{figure:package} shows a UML package diagram of the framework. The packages in the UML diagram correspond to the Java packages in the source code. Please note that the UML diagrams in this thesis are often simplified and do not cover all details.

\subsubsection{Driver}

The driver package is responsible for the low level communication with the block device over USB. It uses USB bulk transfers to access and to communicate with the USB device. It contains the SCSI commands described in the theory part (section \ref{section:scsi_commands}). The package provides methods for reading and writing raw data from and to the device storage.

\subsubsection{Partition}

This package is relevant for handling partition tables and recognizing the different partitions and their file systems on a mass storage device. Thus it needs direct access to the block device, as well as access to the file system implementations. The partition package contains code for handling MBR partition tables.

\subsubsection{File system}

The fs package contains the code for the FAT32 file system. It needs direct access to the block device's raw data, in particular to the raw data of the specific partition it represents. That means it only has indirect access to the driver package. All method calls to read or write raw data are routed through the partition package, to handle the different partitions on a block device correctly.

\section{Using the Framework}

The UsbMassStorageDevice class is the main entry point for accessing mass storage devices. It provides a static method which returns all available mass storage devices. This method loops through all connected USB devices and checks if the connected device is a valid device following the USB mass storage class. The mass storage devices can then be initialized, via the init() method. The initialization process consists of reading the partition table, creating the corresponding partitions and evaluating the desired file system for each partition.  The available partitions can then be accessed easily via a getter. The close method closes the USB communication and releases the USB Interface.

\begin{figure}[h!]
\caption{UML class diagram of the UsbMassStorageDevice}
\centering
\includegraphics[scale=0.85]{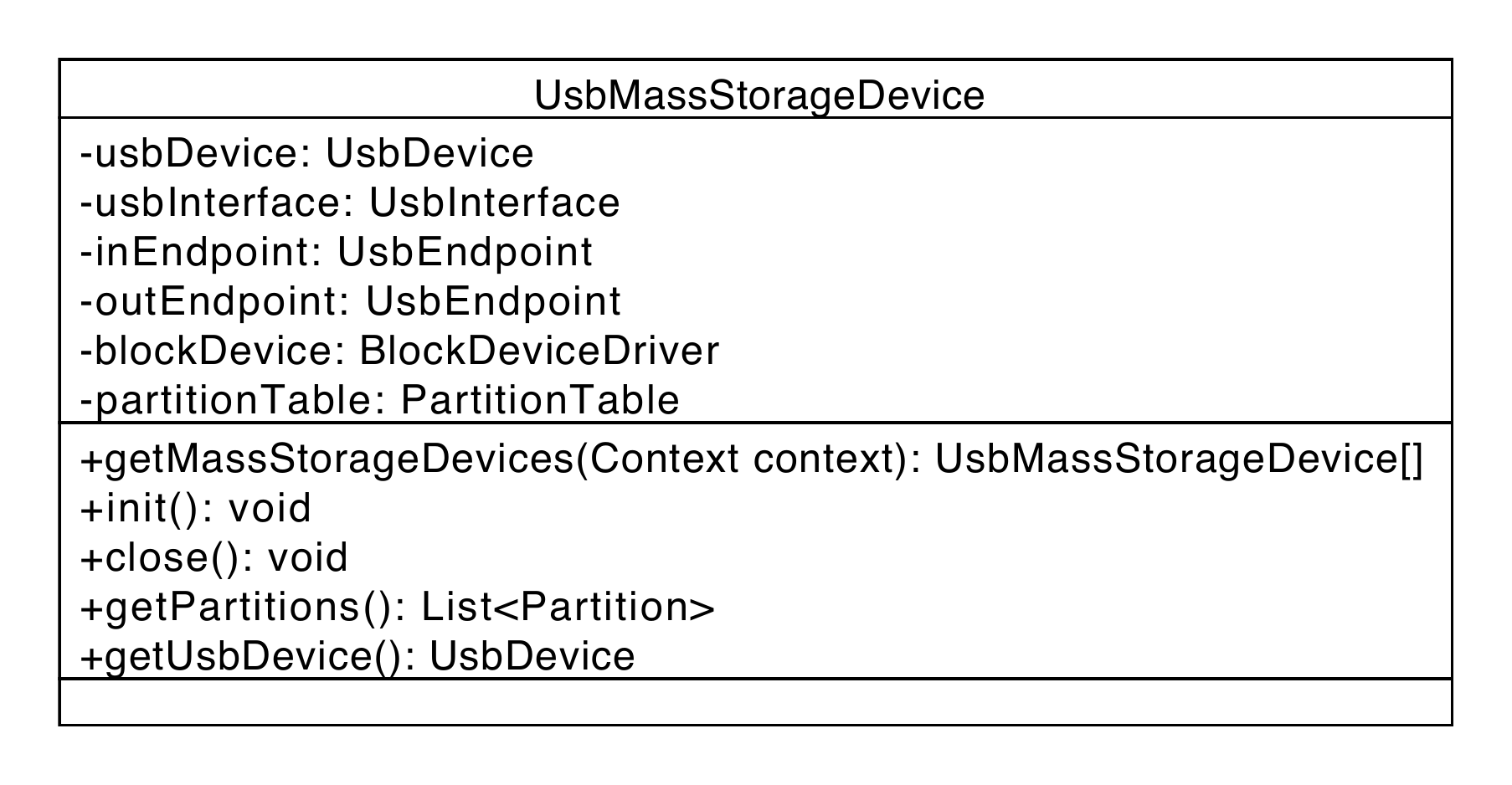}
\label{figure:mass_dev}
\end{figure}

The class also has a lot of private members for communicating with the USB device via the Android API. There is a getter for the underlying UsbDevice, mainly for requesting the permission for communication by the user. Requesting permission is described in chapter \ref{chapter:usb_on_android}, which is about the Android USB host API.

\subsubsection{Partition}

The class Partition represents a single volume on a mass storage device. It provides a getter for the volume label and a getter for the file system to access the contents of the partition.

\subsubsection{FileSystem}

The FileSystem interface provides a getter for the volume label, which returns exactly the same string like the getter in the class Partition. In fact the getter of the Partition class simply delegates the call to the FileSystem class. Another important method allows accessing the root directory of the file system.

\subsubsection{UsbFile}

The UsbFile interface represents an abstraction for files and directories. Every directory or file is an UsbFile. The root directory returned by the FileSystem interface is also a UsbFile. The UsbFile interface provides various methods to access and modify the contents of a file or directory. A complete documentation on every method offers the javadoc in the source code. Note that some methods only make sense for directory or files, but not for both of them!

\begin{figure}[h!]
\caption{UML class diagram of the UsbFile interface}
\centering
\includegraphics[scale=0.85]{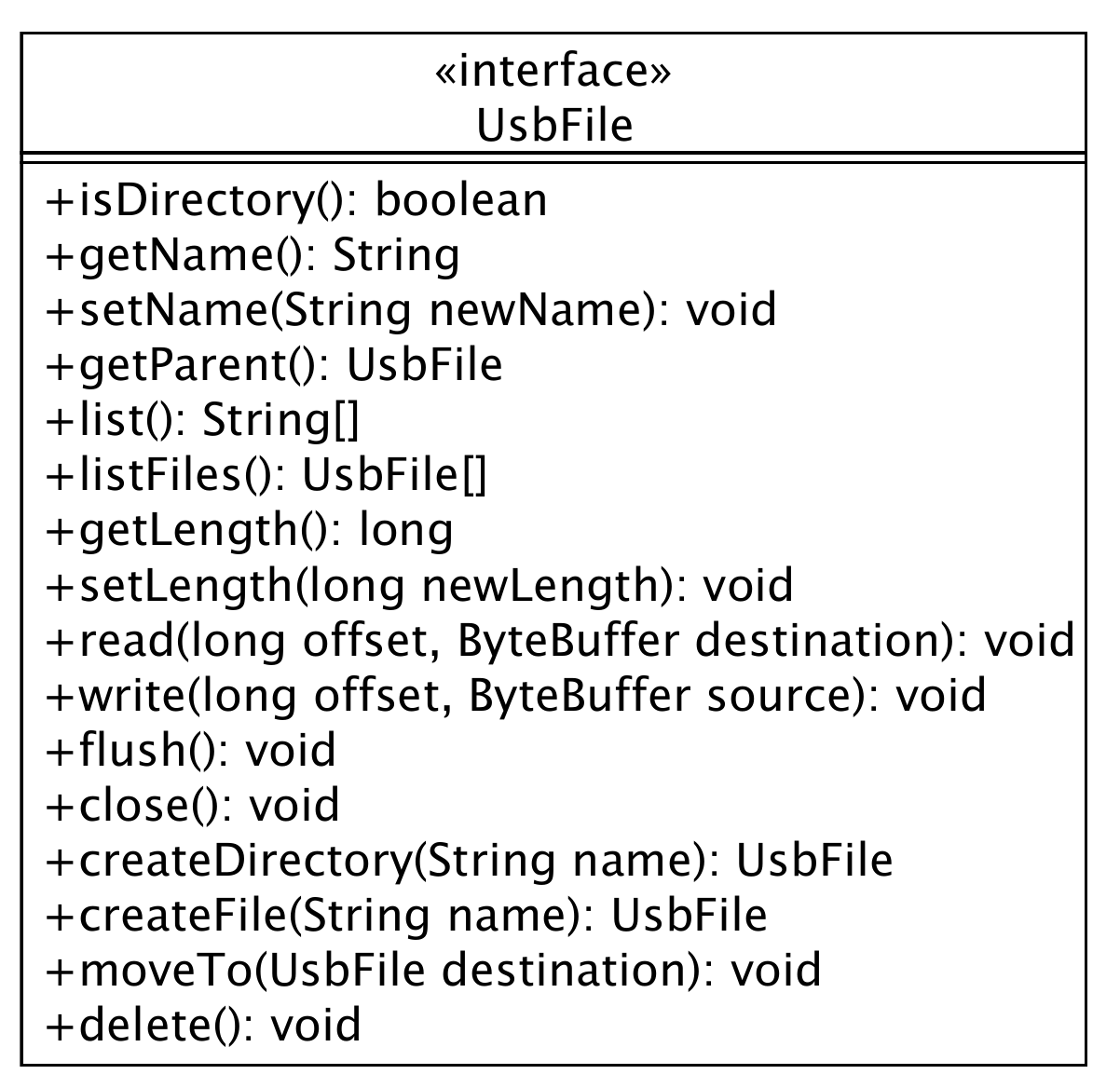}
\label{figure:usb_file}
\end{figure}

\subsubsection{Code example}

The following code demonstrates the use of the classes introduced above. The example simply takes the first USB mass storage device which was found and lists the contents of the root directory of the first partition.

\lstset{language=Java}
\begin{lstlisting}[caption=Code example for accessing the contents of a mass storage device, label=listing:main_example]

private void setupDevice() {
    // the getter needs a Context (Activity or Service) as parameter
    UsbMassStorageDevice[] devices = UsbMassStorageDevice.getMassStorageDevices(this);
    		
    if(devices.length == 0) {
        Log.w(TAG, "no device found!");
        return;
    }
	
    UsbMassStorageDevice device = devices[0];
	
    try {
        // before initializing the device, user must grant permission to communicate
        // this can be done with the UsbManager class and a BroadcastReceiver like shown in the
        // section about the Android USB host API
        device.init();
		
        // always use the first partition of the device
        FileSystem fs = device.getPartitions().get(0).getFileSystem();
        Log.d(TAG, "volume label: " + fs.getVolumeLabel());
		
        UsbFile root = fs.getRootDirectory();
        String[] contents = root.list();
        for(String str : contents) {
            Log.d(TAG, str);
        }
    } catch (IOException e) {
        Log.e(TAG, "error setting up device", e);
    }
}
\end{lstlisting}

\chapter{Inside the packages}

\section{The driver package}

As mentioned previously, the driver package is responsible for handling the low level communication with the block device. It can access the USB device via bulk IN and OUT transfers. Currently only a block device driver for the SCSI transparent command set is available.

\begin{figure}[h!]
\caption{UML class diagram of the driver package}
\centering
\includegraphics[scale=0.8]{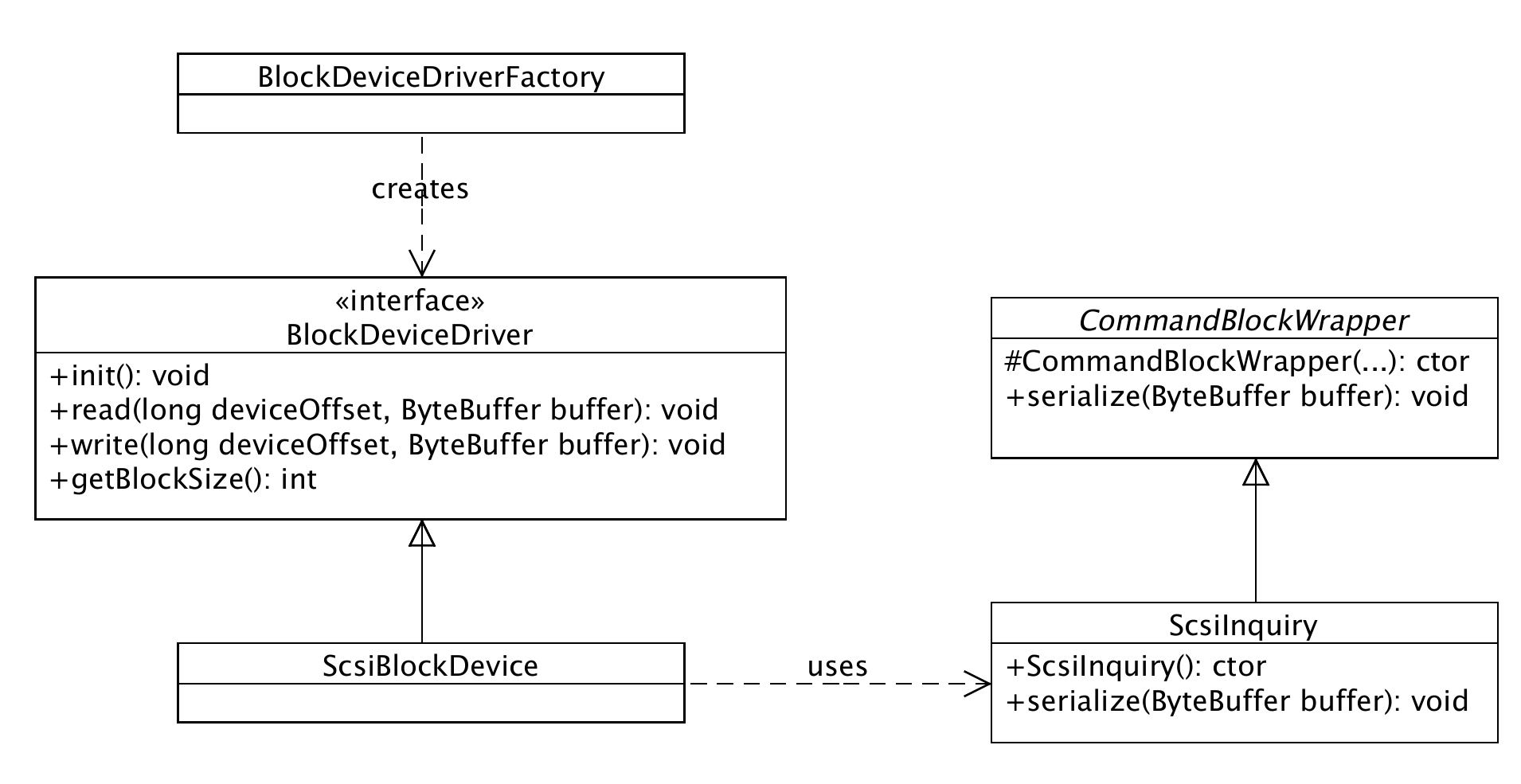}
\label{figure:driver_package}
\end{figure}

\subsubsection{BlockDeviceDriver}

The BlockDeviceDriver interface is a general representation of a block device. It provides methods for reading and writing raw data from and to the device's media storage. It takes a device offset and a ByteBuffer as parameter to determine the offset of a read or write. The ByteBuffer indicates the length of the data which shall be read or written. If data shall be read, the data is read into the ByteBuffer, otherwise the data in the ByteBuffer is written to the device. It also offers a getter to determine the block size of the connected device.

\subsubsection{BlockDeviceDriverFactory}

This class is in charge of creating a suitable block device driver for the connected mass storage device. Currently it always creates a ScsiBlockDevice. No other driver is currently supported. This class is intended to make further development and integration of other device drivers easier. There are also factory classes in the two other packages for creating suitable partition tables and file systems.

\subsubsection{ScsiBlockDevice}

This is the representation of a block device driver which uses the SCSI transparent command set for communicating with devices. It transfers SCSI commands to the device, receives the desired responses from the device and interprets them.

All SCSI commands a modeled in an own class which extend the CommandBlockWrapper. The CommandBlockWrapper is an abstract class which is always coupled with a SCSI command. As already mentioned, every SCSI command is enclosed by a CBW in the SCSI transparent command set protocol. The CBW offers a method to serialize itself to a ByteBuffer. This data can then be transmitted directly to the device. The serialized data include the direction of the command, the transfer length in the transport phase and the length of the SCSI command.

Every SCSI command also offers the serialization to a ByteBuffer, it first calls the serialization method of the CBW class (super class) and then adds the own data to the ByteBuffer. Using this approach it is easy to wrap the CBW around the SCSI commands and new commands are straightforward to implement. 

In the UML diagram \ref{figure:driver_package} only the SCSI INQUIRY command is shown, but there are, of course, also classes for all other commands presented earlier.

\section{The partition package}

\begin{figure}[h!]
\caption{UML class diagram of the partition package}
\centering
\includegraphics[scale=0.8]{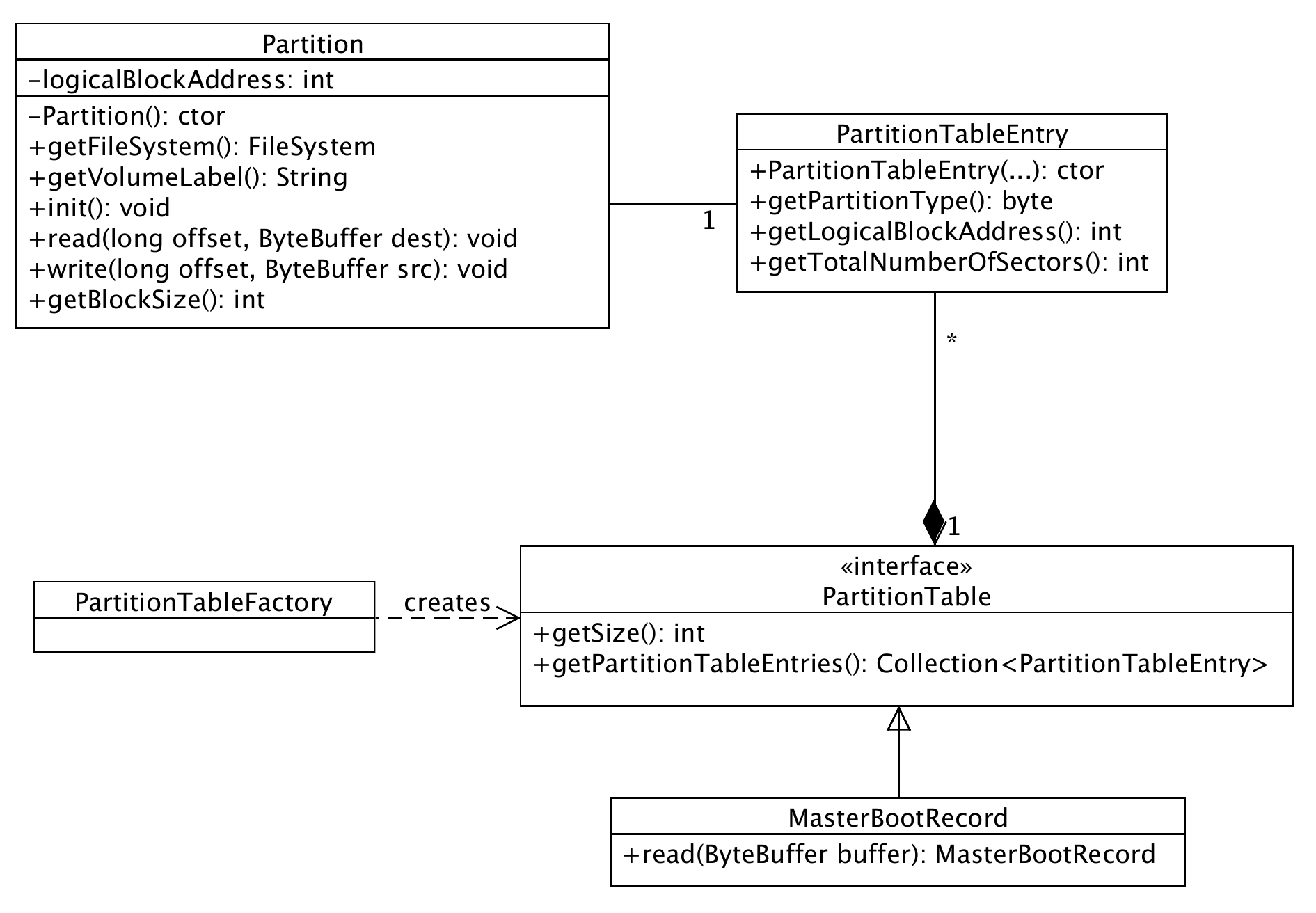}
\label{figure:partition_package}
\end{figure}

The partition package is responsible for handling the partition table on a USB mass storage device. Currently only the MBR partition table is supported. Determining the partition table can be pretty hard, because there is no hint which type of partition table is stored on the device. Therefore the data at LBA zero has to be read (normally a partition table starts at the beginning of a volume) and it has to be checked if the data represents a valid partition table. Figure \ref{figure:partition_package} illustrates the contents of the partition package. There is again a factory class for creating suitable partition tables.

\subsubsection{PartitionTable}

This interface represents in general a partition table. It provides a getter to receive all partition table entries in the table. There is also a method for getting the size of the partition table. For the MBR this is 512 bytes. The factory class uses this size, to determine how many bytes from the mass storage device have to be read.

\subsubsection{PartitionTableEntry}

The PartitionTableEntry represents the information of a partition stored in the partition table. It saves the logical block address where the partition starts, the total number of sectors/blocks the partition occupies and the type of the partition. This is mostly the file system type of the partition, but in case of the MBR this can also be an extended partition.

\subsubsection{MasterBootRecord}

This class covers the Master Boot Record implementation. It has a static read method which returns an instance of the MasterBootRecord class, or null if the data in the ByteBuffer does not look like a Master Boot Record.

\subsubsection{Partition}

The Partition class was already introduced in the overview section (\ref{implementation_overview}), but this time the focus lies on the interaction with other classes of the package and the file system package. 

The Partition class has access to the PartitionTableEntry it represents. It uses the information stored in the entry to determine the starting point (LBA) of the partition and to initialize a suitable file system, the user can then access. The Partition class implements also the BlockDeviceDriver interface from the driver package, described earlier. This is needed because every partition represents an independent volume of the mass storage device. A file system driver should be independent of the location of a partition, hence the partition is responsible for translating the requests of the file system driver according to the starting point, the logical block address, of the partition.

\section{The file system package}

The main classes to access the file system and the contents of it, are the previously mentioned FileSystem class, and the UsbFile interface. These two classes are not explained again, instead the focus lies on a deeper insight into the implementation of the FAT32 file system.

\begin{figure}[h!]
\caption{UML class diagram of the FAT32 implementation}
\centering
\includegraphics[scale=0.9352]{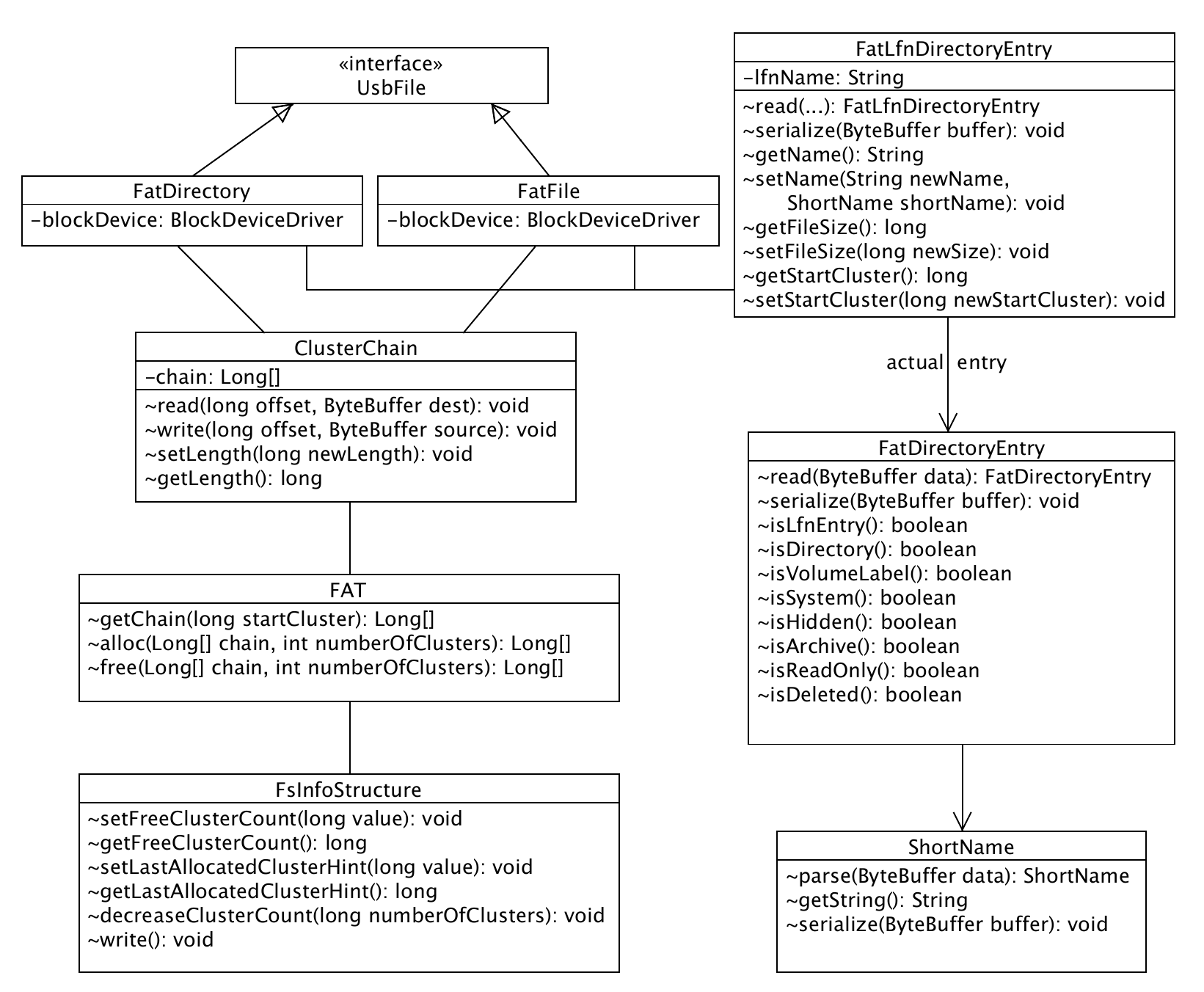}
\label{figure:fat_package}
\end{figure}

\subsubsection{FatDirectory and FatFile}

The FatDirectory and FatFile classes are responsible for the directory and file handling. They have two important attributes, the ClusterChain and the FatLfnDirectoryEntry. The FatLfnDirectoryEntry contains the long file name of the directory or file, the start cluster and if it is a file, the length of a file. It also contains information about the date and time the directory or file was created, last accessed and modified. 
The ClusterChain is responsible for accessing the data located on the disk. A FatDirectory parses and writes the directory entries located in the chain, while the FatFile passes the functionality of reading or writing directly to the user via the UsbFile interface.

\subsubsection{FatLfnDirectory and FatDirectoryEntry}

The FatLfnDirectory provides methods to access a long file name entry. It encapsulates the long file name as well as the actual entry holding important information about the entry, such like the start cluster or file size, and the short name. The FatLfnDirectory entry mostly delegates the calls which do not have to do with the long file name, such as the getting or setting the start cluster or file size, to the actual entry which is an attribute of the FatLfnDirectory class. Both classes and the ShortName class provide convenience methods for parsing/reading and serializing the data. 

\subsubsection{ClusterChain and FAT}

The ClusterChain handles reading and writing from and to cluster chains given a certain start cluster. It offers read and write methods to read and write the raw data from and to the desired clusters on the disk. It also has methods to set and get the length of the cluster chain. The setLength() method, grows and shrinks the cluster chain as needed with the help of the FAT class. When writing to the chain, the cluster chain will also be dynamically increased if needed.

The FAT is responsible for the cluster distribution in the File Allocation Table. The method getChain() returns an array containing all clusters in the chain, including the start cluster at the first position, given a certain start cluster. The FAT class can also allocate new clusters or free unneeded clusters from a chain. When allocating or freeing clusters the FAT class also sets the new information regarding the free clusters and the last allocated cluster hint in the FSInfoStructure.

\subsubsection{Fat32FileSystem and Fat32BootSector}

These two classes are not shown in the UML class diagram \ref{figure:fat_package}, because they do not play an important role in the interaction of the classes. But nevertheless, without them the whole system would not work. The Fat32FileSystem class is mainly responsible for initializing the file system, meaning reading the boot sector, preparing the FAT and FSInfoStructure. It is also responsible for initializing the root directory and to hand it (via a getter) to the user, if desired.

The Fat32BootSector class reads the information of the boot sector of the FAT32 file system, and provides getter for important information, other classes need to access. This information includes but is not limited to the cluster and sector size and the start cluster of the root directory.

%% file: chapters/quality_management.tex
\chapter{Testing}

\section{Overview}

To ensure the quality of the Framework it has been tested on a wide range of different Android devices. The framework has also been tested with different USB pen drives and an external HDD, with external power source. Card readers have not been tested! The results of the tests are explained in the following sections.

\subsection{Testing criteria}

On every device following aspects were tested, if they succeeded or not:

\begin{itemize}
\item Listing the contents of directories
\item Reading and writing from and to files
\item Adding directories and files
\item Removing directories and files
\item Renaming directories and files
\item Moving directories and files to other directories
\item Writing files bigger than the cluster size, to check if the dynamic growing of a cluster chain works correctly
\end{itemize}

\section{Results}

Table \ref{table:test_results} shows the test result for the devices which have been tested. If all aspects work properly, the test is successful.

\begin{table}[ht]
\caption{Test results}
\centering
\begin{tabular}{|l|l|l|p{5.5cm}|}
\hline\hline
\textbf{Device} & \textbf{Android Version} & \textbf{Success} & \textbf{Comments} \\ \hline
Archos 101 G9 & 4.0.4 & Yes & Has native support for USB mass storage devices. \\ \hline
Google Nexus 4 & 4.4.2 & No & Does not have the USB host feature\cite{nexus_4_usb_host}. \\ \hline
Google Nexus 5 & 4.4 & Yes & - \\ \hline
Google Nexus 7 & 4.2.2 & Yes & - \\ \hline
Google Nexus 7 & 4.4.2 & Yes & - \\ \hline
Google Galaxy Nexus & 4.3 & Yes & - \\ \hline
Google Nexus S & 4.1.2 & No & Does not have the USB host feature. \\ \hline
Samsung Galaxy S3 & 4.3 & Yes & Has native support for USB mass storage devices. \\ \hline
\end{tabular}
\label{table:test_results}
\end{table}

\subsection{Native support}

Some devices support mounting USB mass storage devices without root rights natively. This was verified for the Samsung Galaxy S3 and the Archos 101 G9. When connecting a mass storage device via the USB OTG adapter to such a device, the mass storage device is automatically mounted and can be either accessed with the file manager which came with the device or any third party file manager. On the Archos device the mass storage is mounted under \textit{/mnt/ext\_storage}.

\subsection{Performance test}

On Android versions lower 4.3 a specific method in the API is not available. It was later added with API level 18. To support older Android versions and to overcome this lack, the framework uses a different API call on lower Android versions. This workaround can influence on the performance. To examine the performance, the same file is copied from the mass storage to the internal storage on different Android versions, but on an identical device. More information about the API difference, can be found in appendix \ref{chapter:api_diff}.

The file copied was a video file with a size of 155,883,762 bytes, which is approximately 148.6 megabytes. The two devices were Google Nexus 7 tablets with Android versions 4.4.2 and 4.2.2. The same USB pen drive was used for the tests. The file was copied five times on every device, table \ref{table:performance_test} shows the average copy time.

\begin{table}[ht]
\caption{Performance test: Average time of copying same file five times on each device}
\centering
\begin{tabular}{|l|l|l|l|}
\hline\hline
\textbf{Android Version} & \textbf{Time in Milliseconds} & \textbf{Time in seconds} & \textbf{Time in minutes} \\ \hline
4.2.2 & 90203.1 & $\sim$ 90 & $\sim$ 1.5 \\ \hline
4.4.2 & 100040.6 & $\sim$ 100 & $\sim$ 1.6 \\ \hline
\end{tabular}
\label{table:performance_test}
\end{table}

The performance results are pretty interesting because the device with the lower Android version is definitely faster than the one with the newer one. The difference is about ten seconds! The devices behave contrary as assumed. The reason for this is hard to find, maybe the USB host stack has changed greatly between these two versions. But this would imply that the newer Android version provides an inferior performance consulting the USB host support. Another reason could be that installed applications and running services in the background have a huge impact on the performance. This is also an evidence for the difficulty of creating reasonable performance tests on two different devices, although if they both are of the same model!

\subsubsection{Additional performance test}

Because of the surprising results another test on the device with Android 4.2.2 was run. The test is exactly as the first test, except that this time, in the first test run the offset to write in the buffer is always zero, the second time it is forced to be none zero. That means that in the second run the workaround is enforced and the buffer has to be copied. Again, more information regarding the API difference can be found in appendix \ref{chapter:api_diff}.

\begin{table}[ht]
\caption{Performance test: Average time of copying same file five times on same device}
\centering
\begin{tabular}{|l|l|l|l|}
\hline\hline
\textbf{Test Run} & \textbf{Time in Milliseconds} & \textbf{Time in seconds} & \textbf{Time in minutes} \\ \hline
Offset zero & 84751 & $\sim$ 85 & $\sim$ 1.4 \\ \hline
Offset non zero & 90203.1 & $\sim$ 90 & $\sim$ 1.5 \\ \hline
\end{tabular}
\label{table:performance_test2}
\end{table}

This time the result is as expected, meaning the overhead of copying the whole data into a temporary buffer takes (about five seconds) longer.

%% file: chapters/results.tex
\chapter{Summary}

The goal of the thesis was to develop a framework, for the Android operating system, which allows access to USB mass storage devices. These devices include USB pen drives, card readers or even external HDDs. The framework allows discovering and exchanging data with these devices in terms of directories and files.

The thesis gives an overview of basic aspects relating USB in general and how to use the Android USB host API. A very important aspect is the description of the USB mass storage class, including the bulk-only transfer and the SCSI transparent command set. The most important SCSI commands were also introduced. After that a detailed view on the theory about file systems, in detail the FAT32 file system, follows.

After the theoretical part, the developed framework is described, it's purpose, general structure and most important aspects in detail.

The part about the quality management gives a short overview of the interplay of the framework with different devices and Android versions.

\section{Current status}

The Framework works on every device with Android 3.1 or later, which has hardware and software enabled USB host support. Every Android device which meets these requirements should be supported. The framework currently supports mass storage devices using the bulk-only transport with the SCSI transparent command set. The mass storage device must be formatted with an MBR, located at the beginning of the storage. There is no support for other partition tables. Devices completely without an partition table are also unsupported. The supported file system is FAT32, which is the most commonly used one for USB mass storage devices. Other file systems, even the one from the FAT family, like FAT12 or FAT16, are not supported.

Despite these limitations the framework has all features which had been determined at the beginning of this thesis\footnote{The desired requirements and features are described in section \ref{implementation_purpose}.}. In the source code, there are currently some TODOs to mark starting points for minor enhancements. But these points do not compromise the everyday use. Currently there are a lot of debug log messages which help to understand the operation of the framework and may decrease the performance, especially when reading or writing huge amounts of data from or to files.

For example, the SCSI REQUEST SENSE command could be added. Further information about an unsuccessful command can be acquired by it. This is not a serious problem, because it is very rare that a device cannot execute a command successfully. During the development, this constellation occurred only when the commands transfered to the device were incorrect and thus the framework had a bug. For the development of extensions to the framework, the addition of this command may be helpful.

Nevertheless, the framework is very easy to extend. Most parts operate independent from each other and can be easily be exchanged. Other block device drivers, partition tables or file systems can easily be added without changing unrelated parts. This is why the framework often relies on interfaces instead of particular implementations and often uses factory classes for the initialization of instances.

\section{Conclusion}

Developing Android apps, is surely much fun. The Android API is well structured, organized and easy to understand. This also stays on when developing advanced applications like this one. 

Very interesting was the aspect of developing low level things like a block device driver and a file system driver in the Java programming language. Normally these things are done in C, and not in a higher level language like Java. But solely relying on Java and strictly avoiding C code was not a problem at any time, Java did the job very well! This shows that Java is also perfectly capable of bringing the object oriented approach to things located at lower levels in an operating system or the kernel.

The documentation on the USB mass storage class and the SCSI transparent command set is very rare and often complex. It needs a lot of foreknowledge on some topics. Nevertheless after consulting various resources and spending a lot of time reading, the comprehension continuously increases.

\chapter{Outlook}

The developed framework is usable perfectly at its current status. As always there is room for further development and features. Integrating other block device drivers and file systems are features that come instantly in mind. Maybe someday someone likes to connect his external CD/DVD drive to his Android device?

But not only other block device and file system drivers are possible. Currently the framework has no intelligent caching mechanisms. However it always writes the changes directly to the storage which may be inefficient sometimes. Reading data, at the moment, is implemented straightforward. The appropriate data is just read from disk, there are no special strategies like reading ahead or guessing what data the user wants to access next. These are all things which are available in every up to date operating system and many people invested a lot of time in efficient strategies for caching, etc. Maybe implementing such techniques increase the user experience.

Another useful extension to the framework could be the integration into the new FileSystem API of Java 7\footnote{\url{http://docs.oracle.com/javase/7/docs/technotes/guides/io/fsp/filesystemprovider.html}}. This would give the user the ability to use the default Java API for accessing files. Another benefit could be the use of pipes to integrate the mass storage device in the internal file system of Android. The framework would then run in background and listen for events relating the piped directories and files. Every change to the piped structure would then be written back to the actual mass storage device. With this solution the mass storage device is mirrored into the Android file system.

Maybe with the next Android version Google adds native support for USB mass storage devices in their stock Android version while making this framework (nearly) obsolete. Some manufacturers already noticed that this feature is pretty useful especially when looking at the latest trend omitting a slot for micro SD-cards and relying only on the internal storage of a device. But we will see what Google brings next!

%% file: chapters/oneAppendix.tex
\chapter{Debugging applications via Wifi}
\label{chapter:DetailedDescriptions}

When working with the USB features Android provides, an application on the device cannot be debugged as usual using a USB cable and plugging it into the computer. Android fortunately provides an easy solution for that. It allows debugging over Wifi just like using a USB cable.

To enable debugging over Wifi certain steps have to be done. The first step is to connect the device as usual to the computer and to execute the following command using adb:

\lstset{language=bash}
\begin{lstlisting}[caption=Restarting the device in Wifi debug mode, label=listing:wifi_debug]
localhost:platform-tools mep$ ./adb tcpip 5555
restarting in TCP mode port: 5555
\end{lstlisting}

This command forces the device to restart the debugging functionalities in Wifi mode at port 5555. The device can now be used to debug over Wifi. To do this, the device's IP address has to be looked up in the Wifi settings\footnote{The computer and the Android device, obviouly, have to be in the same (Wifi-)network.}. With the IP address the connection can easily be established with another adb command:

 \begin{lstlisting}[caption=Connecting to the device over Wifi, label=listing:wifi_connect]
localhost:platform-tools mep$ ./adb connect 192.168.2.108
connected to 192.168.2.108:5555
 \end{lstlisting}
 
 After that, deploying and debugging applications can be done, just as usual, in eclipse or the desired environment!

\chapter{Isochronous USB transfers}

The introduction says that Android currently does not support isochronous USB transfers, and gives a reference to the official Android developer documentation\cite{android_usb_constants}. But this does not seem to be appropriate for every device.

Some devices like the Samsung Galaxy S3 support audio output via a connected USB audio interface. Audio input seems to be unsupported. This feature is for example useful for USB headsets or docking stations which can play music. Unfortunately it is not part of the official Android. Nevertheless an application developer has no access to the isochronous transfers, they are hidden in the system and are only used to route the operating systems audio through the connected audio interface instead through the internal speaker.

The company beyerdynamic offers a headphone with an integrated amplifier and an USB audio interface. It is connected to the Android device over USB. The headphone can be used with some Android devices only, which support digital audio data via USB, for example the Samsung Galaxy S3 or S4, or the HTC Butterfly\cite{beyerdynamic}.

In fact, it seems that there are also people who get isochronous transfers to work on non rooted devices using native code and the system call \textit{ioctl}\cite{usb_audio_recorder, iso_stack}.

\chapter{API difference}
\label{chapter:api_diff}

The section about the performance test stated that there is an API change between Android API level 18 and lower which affects the framework. In Android API level 18 the class UsbDeviceConnection offers two methods for performing bulk transfers\cite{android_usb_dev_con}:

\lstset{language=Java}
\begin{lstlisting}[caption=Bulk transfers in UsbDeviceConnection, label=listing:wifi_connectg]
int bulkTransfer(UsbEndpoint endpoint, byte[] buffer, int offset, int length, int timeout);
int bulkTransfer(UsbEndpoint endpoint, byte[] buffer, int length, int timeout);
\end{lstlisting}

One method accepts an offset. It represents the index of the first byte where a read or write shall begin in the byte array. The framework developed in this thesis makes intense usage of the ability to specify the offset. The other method just begins reading or writing at offset zero in the byte array.

In API level lower 18 only the latter method is available. The only workaround which solves this issue is to create a temporary buffer, perform the bulk read or write and then copy the temporary buffer to the actual buffer at the desired offset. This yields to the overhead of an extra array copy. Listing \ref{listing:workaround_api} shows how this is solved in the framework. More information can be found in the source code, especially in the class \mbox{UsbMassStorageDevice}.

\begin{lstlisting}[caption=Workaround for the missing method in API level lower 18, label=listing:workaround_api]
@Override
public int bulkOutTransfer(byte[] buffer, int offset, int length) {
	if(offset == 0)
		return deviceConnection.bulkTransfer(outEndpoint, buffer, length, TRANSFER_TIMEOUT);
	
	byte[] tmpBuffer = new byte[length];
	System.arraycopy(buffer, offset, tmpBuffer, 0, length);
	int result = deviceConnection.bulkTransfer(outEndpoint, tmpBuffer, length, TRANSFER_TIMEOUT);
	return result;
}

@Override
public int bulkInTransfer(byte[] buffer, int offset, int length) {
	if(offset == 0)
		return deviceConnection.bulkTransfer(inEndpoint, buffer, length, TRANSFER_TIMEOUT);
	
	byte[] tmpBuffer = new byte[length];
	int result = deviceConnection.bulkTransfer(inEndpoint, tmpBuffer, length, TRANSFER_TIMEOUT);
	System.arraycopy(tmpBuffer, 0, buffer, offset, length);
	return result;
}
\end{lstlisting}